\documentclass[preprint, 12pt]{elsarticle}

\usepackage{aas_macros}

\journal{Computational Physics}

\usepackage[utf8]{inputenc}
\usepackage{natbib}
\usepackage{amsmath}
\usepackage{amssymb}
\usepackage{lmodern}
\usepackage{graphicx}
\usepackage{color}

\long\def\beginpgfgraphicnamed#1#2\endpgfgraphicnamed{\includegraphics{#1}}

\begin{document}

\begin{frontmatter}

\title {A Simple Multigrid Scheme for Solving the Poisson Equation with Arbitrary Domain Boundaries.}

\author[cea]{Thomas~Guillet}
\author[cea,zurich]{Romain~Teyssier}
\address[cea]{IRFU/SAp, CEA Saclay, Bâtiment 709, 91191 Gif-sur-Yvette Cedex, France}
\address[zurich]{Institute of Theoretical Physics, University of Zürich, Winterthurerstrasse 190, 8057 Zürich, Switzerland}

\begin{abstract}
We present a new multigrid scheme for solving the Poisson equation with Dirichlet boundary 
conditions on a Cartesian grid with irregular domain boundaries. This scheme was developed in the context of the
Adaptive Mesh Refinement (AMR) schemes based on a graded-octree data structure. The Poisson equation
is solved on a level-by-level basis, using a ``one-way interface'' scheme in which boundary conditions are interpolated from the previous 
coarser level solution. Such a scheme is particularly well suited for self-gravitating astrophysical flows requiring an adaptive time stepping strategy.
By constructing a multigrid hierarchy covering the active cells of each AMR level, we have designed a memory-efficient algorithm that can benefit fully 
from the multigrid acceleration.
We present a simple method for capturing the boundary conditions across the multigrid hierarchy, based on a second-order accurate reconstruction of the boundaries of the multigrid levels.
In case of very complex boundaries, small scale features become smaller than the discretization cell size of coarse multigrid levels and convergence problems arise. We propose a simple solution to address these issues.
Using our scheme, the convergence rate usually depends on the grid size for complex grids, but good linear convergence is maintained.
The proposed method was successfully implemented on distributed memory architectures in the RAMSES code, for which we present and discuss convergence and accuracy properties as well as timing performances.
\end{abstract}


\end{frontmatter}

\newcommand{\bigO}{\mathcal{O}}
\newcommand{\assign}{\longleftarrow}

\section{Introduction}

Elliptical partial differential equations play a central role in many
mathematical and physical problems. The Poisson equation, in
particular, arises naturally in the context of electromagnetism, fluid
dynamics and gravity. It is therefore of great significance in
astrophysics, of which those fields are essential theoretical
aspects. In the classical Newtonian limit, one can relate the
gravitational potential $\Phi$ to the matter distribution (local
density) $\rho$ by:
\begin{equation}
 \Delta \Phi = \rho
 \label{eq:poisson}
\end{equation}
Applications in computational astrophysics therefore often require
solving a discretization of Eq. (\ref{eq:poisson}) on some discrete
grid. The ubiquity of the Poisson equation in science has led to the
development of many suitable numerical solving techniques, for a wide
range of specific applications and problems.

% A direct method: FFT
In some cases, one can solve for the discretized potential directly
from Eq.  (\ref{eq:poisson}): the convolution theorem allows to solve
the Poisson equation directly in Fourier space using the Green
function of the Laplacian operator on the grid of interest. In
computational applications, this is usually done using the Fast
Fourier Transform (FFT), which is particularly suited to periodic
boundary conditions problems on rectangular, regular grids.  This
direct method is very effective: on a grid of linear size $N$, the
computing time in 3 dimensions is of order $\bigO(N^3 \log N)$, and
yields the solution to the discretized problem within numerical
roundoff accuracy. The Green function method can be generalized to
non-periodic boundary conditions and adaptive grids on rectangular
domains \citep[see e.g.][]{Huang2000}.

% Iterative methods:
% Relaxation methods
Another class of Poisson solving schemes are relaxation methods, such
as Gauss-Seidel or successive over-relaxation
(SOR) \citep[see e.g.][]{Press1992}. Unlike the Green function method, those schemes are
iterative, and rely on an initial estimate (``first guess'') of the
solution. The estimate is then successively improved by iterative
damping of the residual $r = \Delta \Phi - \rho$.  Relaxation methods
are generally simple to implement, and suitable for a wide range of
elliptic problems. Methods such as Gauss-Seidel or SOR only require
the forward computation of the differential operator, which is usually
inexpensive, and makes it possible to use such relaxation methods on
complex grid geometries.
% Krylov subspace methods
Krylov subspace methods \citep[see e.g.][]{Saad2003}, such as the Conjugate Gradient (CG) method,
are another class of iterative methods which can be used to solve Eq.
(\ref{eq:poisson}) in the form of the optimization problem:
\begin{equation}
 \min_\Phi \left\| \Delta \Phi - \rho \right\|^2
\end{equation}

Classical iterative methods have gained in interest in the last decade
with the advent of massively parallel computing. They only require
one-cell thick, boundary layer from the neighboring processor,
while FFT involves a massive transpose of the whole volume of data.
Moreover, one does not need the exact solution of the discretized Poisson
equation that the FFT provides: the error in the solution is already
dominated by truncation errors of a fraction of a percent for practical
applications. Iterative methods can therefore be stopped quite
early, when the residual norm has dropped well below
the truncation errors. Still, classical iterative methods can be very
time consuming, especially when the grid size is increased. Usually,
the large scale modes are the slowest to converge, and the number
of iterations required to reach a given level of residual usually
increases with the number of grid points per dimension \citep{Press1992}.

The convergence rate of iterative relaxation methods, such as
Gauss-Seidel, can be dramatically improved by multigrid acceleration
\citep{Brandt1977, Wesseling1992}.  Multigrid methods use a hierarchy
of discretizations of decreasing spatial resolution. At the resolution
of the initial problem (on the finest grid), the solution is
iteratively improved by subtracting corrections obtained from the
coarser grids. This ensures that the large-scale modes of the residual
are efficiently damped, while a traditional smoothing relaxation scheme takes
care of the small-scale modes of the error.  In ideal cases (e.g. for
smooth problems with simple boundary conditions on rectangular
domains), multigrid methods can exhibit remarkable convergence
properties: the residual norm decreases at a constant rate during the
whole convergence process (down to machine precision). The convergence
rate can be very fast (see examples below), and more importantly {\it
  it does not depend on the size of the grid}. This last property is
perhaps the most important one, since it allows one to consider very
large computational problems: for a given level of convergence in the
residual norm, optimal multigrid methods are linear in the problem size $N^3$.  This
complexity is lower than the one offered by the FFT, and although
multigrid is still slower than optimized FFT for small configurations, it can be
quite competitive for large parallel computations and reasonable
stopping criteria.  In contrast to traditional iterative schemes, the
convergence properties of multigrid are unaffected by the choice and
quality of the first guess. Another reason to use multigrid is that it
couples nicely to adaptive Cartesian grids, while FFT-based methods in
this case are still possible, but quite complicated \citep[see
  e.g.][]{Cheng1999, Huang2000, Ricker2008}.  Last but not least,
while FFT requires a Poisson equation with constant coefficients,
multigrid methods, because they are defined in real space, can be adapted to solve
Poisson equations with non-uniform coefficients, or even non-linear equations. This
property is particularly useful for problems with complex boundaries
\citep{Popinet2003}, multiple incompressible fluids \citep{Rudman1998}, or models
with modified gravity \citep{Knebe2004,Tiret2007,Llinares2009}).

The objective of this paper is to implement a simple and
efficient multigrid Poisson solver for a Cartesian grid with {\it
  arbitrary domain boundaries}. Complex boundary conditions are
usually found in fluid mechanics problems featuring immersed bodies
or complex multiphase flows \citep{Sussman1994, Sussman1999, Liu2000}. Here, our
motivation is different: we would like to design a Poisson solver for
the class of Adaptive Mesh Refinement (AMR) codes for which the mesh
is defined on a ``graded octree'' \citep{Kravtsov1997,
  Teyssier2002, Popinet2003}.  The octree mesh structure ensures
that the geometry of the grid closely adapts to the properties of the
flow, without the traditional overhead associated to the large
rectangular patches of block-structured AMR. The consequence is
however that the mesh of a given AMR level can have a very complicated
shape, with holes and irregular boundaries. If one solves the Poisson
equation on the whole AMR hierarchy, this translates into a
modification of the Laplacian operator at fine-coarse grid
boundaries, but the corresponding multigrid solver remains similar to
its Cartesian grid equivalent \citep[see e.g.][]{Johansen1998}.

In most astrophysical applications, it is however almost always impractical to
solve the Poisson equation on the whole grid at once: self-gravity has the
effect of dramatically increasing the dynamical range of the density
field, leading to very different characteristic timescales within the
same computational domain. Advancing the whole system with one single
time step can be very inefficient, since only a very small fraction of
the volume actually needs high accuracy in the time coordinate.  This
is particularly true for cosmological simulations, where most of the
computational volume is covered by low density regions that evolve
slowly, while a small number of highly resolved cells sample dense
regions (such as galaxies), where the dynamical time scale is very
short.  Most AMR codes address this problem by using \emph{adaptive
  time stepping}, where a given fine level is updated more frequently than
its coarser level, ensuring that the actual timestep remains close
to the natural CFL timestep (so that the whole AMR grid is not updated
more often than needed).
In RAMSES, we typically update a finer level twice as often as its coarser level.
As a consequence, consecutive fine and coarse levels are only synchronized at
every other fine timestep.

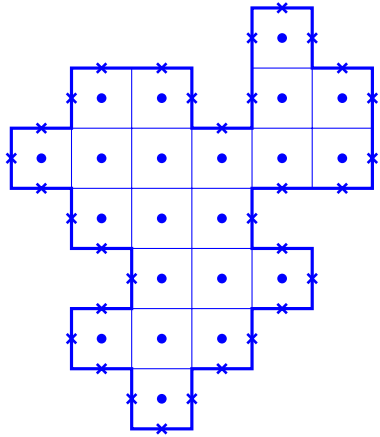
\begin{figure} % {{{ Fig: Fine problem
\begin{center}
\beginpgfgraphicnamed{grid}
\begin{tikzpicture}[scale=0.8, yscale=-1]

  % Define the paths
  \newcommand{\GridContour}{
(1,0)--(1,1)--(0,1)--(0,2)--(1,2)--(1,3)--(2,3)--(2,4)--(1,4)--(1,5)--(2,5)--(2,6)--(3,6)--(3,5)--(4,5)--(4,4)--(5,4)--(5,3)--(4,3)--(4,2)--(6,2)--(6,0)--(5,0)--(5,-1)--(4,-1)--(4,1)--(3,1)--(3,0)--cycle }

  \begin{scope}

  % Draw the boundary condition markers
  \foreach \pos in {
(1,0.5), (0.5,1), (0,1.5), (0.5,2), (1,2.5), (1.5,3), (2,3.5), (1.5,4), (1,4.5), (1.5,5), (2,5.5), (2.5,6), (3,5.5), (3.5,5), (4,4.5), (4.5,4), (5,3.5), (4.5,3), (4,2.5), (4.5,2), (5.5,2), (6,1.5), (6,0.5), (5.5,0), (5,-0.5), (4.5,-1), (4,-0.5), (4,0.5), (3.5,1), (3,0.5), (2.5,0), (1.5,0) }
	\node[boundary point] at \pos {};
  \clip \GridContour;

  % Draw the grid
  \draw[fine cells] (-1,-2) grid (8,8);

  % Draw the cell centers
  \foreach \x in {-1, ..., 10}
    \foreach \y in {-1, ..., 10}
	  \node[fine cell center] at (\x+0.5, \y+0.5) {};

  \end{scope}
  \draw[fine boundary] \GridContour;

\end{tikzpicture}
\endpgfgraphicnamed
\end{center}
\caption{Poisson problem at the fine level for a given AMR level. The solution potential to be determined is defined at the cell centers (dots). The boundary conditions are defined at the border cell faces (crosses). The boundary values are linearly interpolated from the solution at the coarser AMR level (not represented), which was previously computed because of the one-way interface scheme.}
\label{fig:grid}
\end{figure}

Whenever all the AMR levels are synchronized, it is possible to solve
the Poisson equation on the whole grid at once. Such a Poisson solver
is called a ``two-way interface'' scheme, as the information is
propagated both from the coarse grids to the finer grids and back: the
coarse grids ``feel'' the effect of the finer grids. Many multigrid
solvers have been successfully implemented in this case
\citep{Johansen1998, Popinet2003, Miniati2007, Ricker2008}.
\citet{Popinet2003} has developed an incompressible flow solver with complex boundary
conditions, using an octree data structure. The solver features a ``half V-cycle''
multigrid schedule, which makes it possible to perform multigrid iterations
within the existing AMR hierarchy, with no additional storage.
This pioneering work, however, requires the use of a single global time step
for the whole AMR grid.
Most of the time, fine and coarse levels are not
synchronized, and it is not possible to solve on the whole grid at
once. It is therefore necessary to resort to a ``one-way interface''
scheme \citep[see][]{Jessop1994, Kravtsov1997, Teyssier2002,
  Miniati2007}. In the traditional ``one-way interface'' approach, the AMR levels are updated separately: the coarsest
level is updated first, then the computed coarse potential is used to
impose boundary conditions on the finer levels, which can then be
solved separately and more frequently.  The gravitational potential on
finer AMR levels is solved by imposing Dirichlet boundary conditions
at the finer level boundary, enforcing potential continuity. Because
we are dealing with an arbitrarily complex octree structure, we must be
able to solve the Poisson equation on a given refinement level with
arbitrarily shaped domains.  The problem we would like to address here
is therefore to solve the Poisson equation using the multigrid method,
with the constraint that Dirichlet boundary conditions are imposed on
the outer cell faces of an arbitrary domain (see Fig. \ref{fig:grid}).
Note that using adaptive time stepping and one-way interface poses an time synchronization problem:
as the finer levels are stepped forward,
the boundary conditions derived from the coarser levels fall out of synchronization
with the time at finer levels.
The problem of non-synchronized AMR levels has been approached in the context of incompressible flows \citep{Martin2000}
and coupled collisional and collisionless systems for cosmological applications \citep{Miniati2007}.
A proposed solution is to update the coarse levels first and obtain the fine boundary conditions
by both time and space interpolation of the coarse values.
However, this requires storing the time derivatives of the potential at every level.
In addition, in order to obtain a fully multilevel solution, the coarse levels should take into account the solution on the finer levels.
Whenever the coarse levels are updated first, it is therefore necessary to correct their values with an estimate of the difference between the single level and multilevel solutions, before the finer levels are updated.
Such an approach has been used for example by \citet{Martin2000,Miniati2007} in order to provide corrected coarse boundary conditions for the time interpolation.
RAMSES uses a different approach: finer levels are updated first, using as boundary conditions values which are not extrapolated in time but fixed at the initial coarse time step.
Once the fine levels have been stepped forward in time, RAMSES updates the matter density in the coarse cells by binning the fine children cells into the coarse level density with a CIC scheme around their center of mass.
The information from the fine evolution thus flows back up to the coarse levels.
This scheme, however, does not guarantee the time accuracy of the time interpolation and correction methods.
Note that the solver presented in this paper applies equally well to any scheme used to derive the boundary conditions,
as it handles each AMR level independently.

As Cartesian grids are much more practical than unstructured meshes,
the accurate description of boundaries immersed in Cartesian grids
have been the focus of much effort, especially in computational fluid
dynamics \citep[see e.g.][]{Almgren1997, Ye1999}.  This approach has
also led to the development of sophisticated Poisson solvers, some relying on
multigrid acceleration \citep{Johansen1998, McCorquodale2001}.

Our goal is to present a simpler but efficient solver, suitable in
particular to the case of tree-based AMR codes with a one-way
interface scheme.  We will apply our new scheme to the AMR code
RAMSES, a self-gravitating fluid dynamics code for astrophysical
applications \citep{Teyssier2002}.  The octree structure is based on
the ``Fully Threaded Tree'' approach proposed by \citet{Khokhlov1998}
for which cells are split into ``octs'' (groups
of 8 cells) on a cell-by-cell basis  \citep{Kravtsov1997, Teyssier2002}. 
The paper is organized as
follows: we first describe our basic multigrid algorithm for the
Poisson equation with Dirichlet boundary conditions. We then detail
how we capture complex boundary conditions at coarse multigrid levels within our framework.
We report a very fundamental issue of our approach (namely
the ``small islands'' problems) and we propose a simple fix to
overcome it, and we finally discuss the performance of our algorithm.

Note that we only address the issue of the boundary reconstruction for coarsened multigrid levels:
for one-way interface AMR applications, at the finest level (i.e., the AMR level to solve), the boundary is usually located along faces of
Cartesian grid cells constituting the fine AMR level, and no cut-cell or level set reconstruction is needed in this case.
However, we believe our method can be adapted to a cut-cell or level set boundary at the fine level,
provided an appropriate fine operator is defined.

\section{A Multigrid algorithm for complex boundary conditions}

Multigrid methods combine a relaxation solver (usually a smoother such
as Gauss-Seidel or Jacobi sweeps) and a multi-resolution approach, to
ensure efficient damping of both small and large scale components of
the residual (see for example \cite{Wesseling1992} for an
introduction).

\begin{figure} % {{{ Fig: MG hierarchy
\begin{center}
\beginpgfgraphicnamed{hierarchy}
\begin{tikzpicture}[scale=0.6]

  % Define the paths
  \newcommand{\levelA}[1]{
(1,#1,0)--(1,#1,1)--(0,#1,1)--(0,#1,2)--(1,#1,2)--(1,#1,3)--(2,#1,3)--(2,#1,4)--(1,#1,4)--(1,#1,5)--(2,#1,5)--(2,#1,6)--(3,#1,6)--(3,#1,5)--(4,#1,5)--(4,#1,4)--(5,#1,4)--(5,#1,3)--(4,#1,3)--(4,#1,2)--(6,#1,2)--(6,#1,0)--(5,#1,0)--(5,#1,-1)--(4,#1,-1)--(4,#1,1)--(3,#1,1)--(3,#1,0)--cycle }
  \newcommand{\levelB}[1]{
(-1,#1,0)--(-1,#1,2)--(1,#1,2)--(1,#1,6)--(5,#1,6)--(5,#1,2)--(7,#1,2)--(7,#1,0)--(5,#1,0)--(5,#1,-2)--(3,#1,-2)--(3,#1,0)--cycle}
  \newcommand{\levelC}[1]{(-1,#1,-2)--(-1,#1,6)--(7,#1,6)--(7,#1,-2)--cycle }
  \newcommand{\levelD}[1]{\levelC{#1}}

  \tikzstyle{levelA}=[fine cells, thin]
  \tikzstyle{levelB}=[coarse cells, thin]
  \tikzstyle{levelC}=[coarse cells, thin, green!50!black]
  \tikzstyle{levelD}=[coarse cells, thin, black]

  {
  \newcommand{\y}{0}
  \begin{scope}
  \clip \levelA{\y} ;
  \foreach \z in {-1,0,...,7}
	\draw[levelA] (-10, \y, \z) -- (10, \y, \z);
  \foreach \x in {-2,-1,...,6}
	\draw[levelA] (\x, \y, -10) -- (\x, \y, 10);
  \end{scope}
  \draw[levelA] \levelA{\y};
  }

  {
  \newcommand{\y}{2.5}
  \begin{scope}
  \clip \levelB{\y} ;
  \foreach \z in {-2,0,...,6}
	\draw[levelB] (-10, \y, \z) -- (10, \y, \z);
  \foreach \x in {-1,1,...,7}
	\draw[levelB] (\x, \y, -10) -- (\x, \y, 10);
  \end{scope}
  \draw[levelB] \levelB{\y};
  }

  {
  \newcommand{\y}{5}
  \begin{scope}
  \clip \levelC{\y} ;
  \foreach \z in {-2,2,...,6}
	\draw[levelC] (-10, \y, \z) -- (10, \y, \z);
  \foreach \x in {-1,3,...,7}
	\draw[levelC] (\x, \y, -10) -- (\x, \y, 10);
  \end{scope}
  \draw[levelC] \levelC{\y};
  }

  {
  \newcommand{\y}{7.5}
  \draw[levelD] \levelD{\y};
  }

\end{tikzpicture}
\endpgfgraphicnamed
\end{center}
\caption{Levels of the multigrid hierarchy for the problem of Fig.~\ref{fig:grid}.}
\label{fig:hierarchy}
\end{figure}
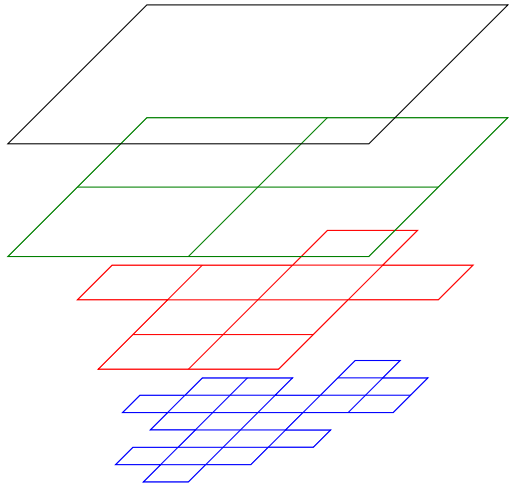

For one-way interface schemes (in the RAMSES code for example), the
Poisson equation is solved over the whole AMR hierarchy on a
level-by-level basis, the size of the cell involved in each individual
Poisson solve being uniform. Each AMR level constitutes the finest
level of the multigrid hierarchy.  The finest multigrid level is
therefore defined by a Cartesian grid with complex irregular
boundaries. We then build a hierarchy of coarser grids which cover
this reference domain (see Fig.~\ref{fig:hierarchy} for an illustration).
This new hierarchy defines the multigrid structure for the current
level. Although it is also based on an octree structure similar to the
underlying AMR grid, it is a different structure that does not
interfere with the coarse AMR levels (which are, in a way,
``orthogonal'' to the multigrid levels).

One major advantage of using this secondary grid hierarchy for each
Poisson solve is that the computational cost of the multigrid solver
at a given level only scales as the number of cells in that level, as
we use only a subset of the AMR coarser cells.  This is especially
crucial for astrophysical problems, where fine grids can occupy a
fraction of the volume much smaller than the coarser levels.

 % }}}

 % }}}

Let us consider the Poisson problem, discretized on a fine grid domain
$\Omega_f$:
\begin{align*}
  \Delta_f \Phi_f = \rho_f \quad & \mbox{on } \Omega_f, \\
  \mbox{Dirichlet conditions} \quad & \mbox{on } \partial \Omega_f.
\end{align*}
where the $f$ subscripts denote the discretized Laplacian operator and
fields on the fine grid.  For a single multigrid iteration, we follow
the traditional recipe:
\begin{enumerate}
  \item Perform $N_{pre}$ Gauss-Seidel smoothing passes on the current estimate
$\Phi_f$ of the fine solution,
  \item Compute the fine residual
	\[ r_f \assign \Delta \Phi_f - \rho, \]
  \item Compute the fine-to-coarse restrictions of the solution and residual:
	\begin{align*}
	  \Phi_c & \assign R(\Phi_f) \\ 
	  r_c & \assign R(r_f),
	\end{align*}
  \item Define a coarsened domain $\Omega_c$ and boundaries $\partial
    \Omega_c$ from $\Omega_f$. The prescription for deriving $\partial
    \Omega_c$ from the fine level is the key issue, and will be
    discussed in section \ref{sec:boundaries}.
  \item Compute a coarse correction $\delta \Phi_c$, by
  performing $N_{cycles}$ multigrid iterations for the following coarse problem:
	\begin{align*}
	  \Delta_c (\delta \Phi_c) = -r_c \quad & \mbox{on } \Omega_c  \\
	  \delta \Phi_c = 0 \quad & \mbox{on } \partial \Omega_c,
	\end{align*}
  \item Prolong $\delta \Phi_c$ into a fine correction
	\[ \delta \Phi_f \assign P(\delta \Phi_c), \]
  \item Correct the fine level solution
	\[ \Phi_f \assign \Phi_f + \delta \Phi_f, \]
  \item Perform $N_{post}$ smoothing passes on $\Phi_f$.
\end{enumerate}
This multigrid iteration is repeated on the fine grid until the norm
of the fine residual is considered small enough. To fully specify the
scheme, one needs to specify integers $N_{pre}$, $N_{post}$ and
$N_{cycles}$, the restriction and prolongation operators $R$ and $P$,
and the discretization of the Laplacian operator.

In our case, we have found $N_{pre}=N_{post}=2$ to yield the best
performance in terms of solver time for our applications in RAMSES.  We will now discuss the
prolongation, restriction and discretized Laplacian operators. We
postpone the discussion of the parameter $N_{cycles}$ to section
\ref{sec:cycles}.

We use a finite difference approach to discretize the Poisson equation
on the Cartesian grid.  Inside the domain, away from the boundaries,
we use the standard 7-point discretization of the Laplacian operator:
\begin{equation}
\left( \Delta \Phi \right)_i = \frac{1}{(\Delta x)^2}\sum_{\{i,j\}} \left( \Phi_j - \Phi_i \right)
\end{equation}
where the sum extends over all the 6 pairs $\{i,j\}$ of neighboring
cells, and $\Delta x$ is the size of the cells on which the Laplacian is
evaluated. This Laplacian operator is second-order accurate and our
multigrid implementation is currently restricted to this case.

For the relaxation smoother, we use a Gauss-Seidel smoother with
red-black ordering, with the Laplacian operator as defined above. With
red-black ordering, the cells are updated in two passes, each pass
running over a half of the domain following the colors of a
checkerboard pattern.

Prolongation and restriction operators define how the problem is
downsampled to a coarser level (restriction) and how the coarser level
correction is interpolated back to the finer level (prolongation).

\begin{figure} %  {{{ Fig: Restriction and interpolation operators
  \begin{center}
  \beginpgfgraphicnamed{restriction}
  \begin{tikzpicture}[scale=0.65]

  \tikzstyle{prolongation}=[->, shorten >=1pt, shorten <=1pt]
  \tikzstyle{restriction}=[prolongation]

  % R1
  \begin{scope}
	\draw node[above] at (3, 6) {$\mathbf R_1$};
	\draw[fine cells] (0,0) grid (6,6);

	\begin{scope}[xshift=1cm, yshift=1cm]
	  \node[fine cell center] (f1) at (1.5,1.5) {};
	  \node[fine cell center] (f2) at (1.5,2.5) {};
	  \node[fine cell center] (f3) at (2.5,1.5) {};
	  \node[fine cell center] (f4) at (2.5,2.5) {};

	  \draw[coarse cells] (1,1) rectangle (3,3);
	  \node[coarse cell center] (c) at (2,2) {};

	  %\draw[restriction] (f1) node[anchor=north east] {$1/4$} -- (c);
	  %\draw[restriction] (f2) node[anchor=south east] {$1/4$} -- (c);
	  %\draw[restriction] (f3) node[anchor=north west] {$1/4$} -- (c);
	  %\draw[restriction] (f4) node[anchor=south west] {$1/4$} -- (c);
	  \foreach \t in {1, ..., 4}
		\draw node[above] at (f\t) {\tiny{1/4}};

	\end{scope}
  \end{scope}

  % R2
  \begin{scope}[xshift=7cm, yshift=0cm]
	\draw node[above] at (3, 6) {$R_2$};
	\draw[fine cells] (0,0) grid (6,6);

	\begin{scope}[xshift=1cm, yshift=1cm]
	  \draw[coarse cells] (1,1) rectangle (3,3);
	  \node[coarse cell center] (c) at (2,2) {};
	  \foreach \x in {0, ..., 3}
		\foreach \y in {0, ..., 3} {
		  \draw node[fine cell center] (f\x\y) at (\x+0.5,\y+0.5) {};
		  %\draw[restriction] (f\x\y) -- (c);
		}

	  \foreach \x in {0, 3}
		\foreach \y in {0, 3}
		  \draw node[above] at (f\x\y) {\tiny{1/64}};

	  \foreach \x in {1, 2}
		\foreach \y in {1, 2}
		  \draw node[above] at (f\x\y) {\tiny{9/64}};

	  \foreach \t in {1, 2} {
		\draw node[above] at (f0\t) {\tiny{3/64}};
		\draw node[above] at (f3\t) {\tiny{3/64}};
		\draw node[above] at (f\t0) {\tiny{3/64}};
		\draw node[above] at (f\t3) {\tiny{3/64}};
	  }
		
	\end{scope}

  \end{scope}

  % P1
  \begin{scope}[xshift=0cm, yshift=-7cm]
	\draw node[above] at (3, 6) {$P_1$};
	\draw[fine cells] (0,0) grid (6,6);
	\node[fine cell center] (f) at (2.5,2.5) {};

	\draw[coarse cells] (1,1) rectangle (3,3);
	\node[coarse cell center] (c) at (2,2) {};

	\draw[prolongation] (c) node[anchor=north east] {$1$} -- (f);
  \end{scope}

  % P2
  \begin{scope}[xshift=7cm, yshift=-7cm]
	\draw node[above] at (3, 6) {$\mathbf P_2$};
	\draw[fine cells] (0,0) grid (6,6);
	\draw[coarse cells] (1,1) rectangle (5,5) (3,1) -- (3,5) (1,3) -- (5,3);

	\node[fine cell center] (f) at (2.5,2.5) {};
	\node[coarse cell center] (c1) at (2,2) {};
	\node[coarse cell center] (c2) at (2,4) {};
	\node[coarse cell center] (c3) at (4,2) {};
	\node[coarse cell center] (c4) at (4,4) {};

	\draw[gray] (f) +(-1, -1) rectangle +(1, 1);

	\draw[prolongation] (c1) node[anchor=north east] {$9/16$} -- (f);
	\draw[prolongation] (c2) node[anchor=south east] {$3/16$} -- (f);
	\draw[prolongation] (c3) node[anchor=north west] {$3/16$} -- (f);
	\draw[prolongation] (c4) node[anchor=south west] {$1/16$} -- (f);
  \end{scope}

  \end{tikzpicture}
  \endpgfgraphicnamed
  \end{center}
  \caption{Common first-order and second-order (respectively left and right) restriction and prolongation schemes (respectively top and bottom). $R_1$ simply averages the values of the subcells to obtain a coarse value. $P_1$ assigns the same coarse value to the all the fine subcells (straight injection).
Our multigrid scheme uses $R_1$ and $P_2$.}
  \label{fig:operators}
\end{figure}
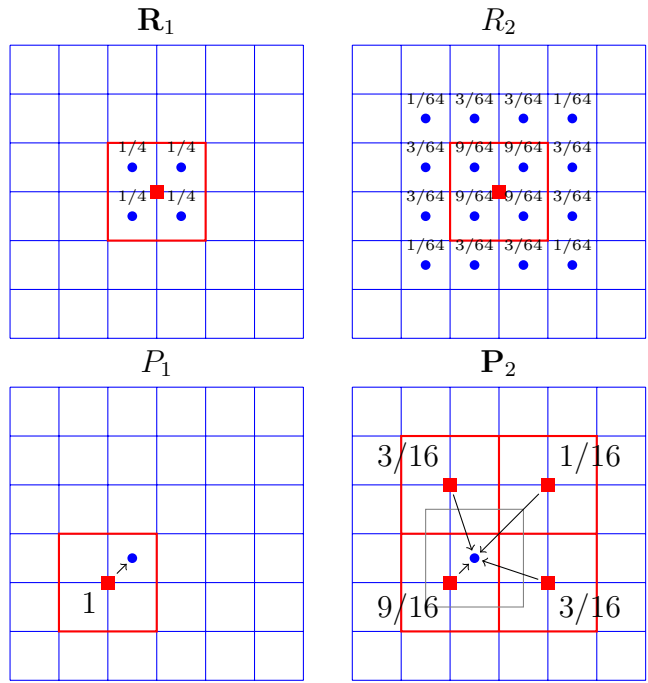 % }}}

A common rule of thumb for the choice of prolongation and restriction operators for the
Poisson equation is the order condition $n_P + n_R > q$, where $n_P$
and $n_R$ are the prolongation and restriction operator orders,
respectively, and $q$ is the order of the Laplacian operator
\citep[see][]{Wesseling1992, Press1992}. Simple standard operators are shown in
Figure~\ref{fig:operators}. Straight injection is defined as $P1$,
while its transpose, $R1$, boils down to a simple averaging. These
operators are first-order in space ($n_R=1$ and
$n_P=1$). Second-order schemes can be easily designed, like
trilinear interpolation (noted here P2) and its transpose (R2), whose
weights are shown in Figure~\ref{fig:operators}.

We have settled for
simple averaging for restriction (R1), and linear interpolation (P2)
for prolongation (see Fig. \ref{fig:operators}), as this choice turned
out to yield the best convergence rates for the classes of grids which
we have studied.  Additionally, as will be discussed in
Section~\ref{sec:Islands}, picking simple averaging as the restriction
operator turns out to be particularly convenient in conjunction with
our prescription for boundary representation.

\section{Second-order multigrid boundary reconstruction and the modified Laplacian operator}
\label{sec:boundaries}

In our multigrid algorithm, the key ingredient is the mathematical
representation of the domain boundary at coarse multigrid levels.
Because our goal is to take advantage of
the Cartesian grid structure of the AMR scheme, we restrict ourselves
to Cartesian grid interface capturing techniques. Among the many
different solutions proposed in the literature, the
cut cell \citep{Johansen1998}
and the level set \citep{Sussman1994, Osher2001, Gibou2002} approaches stand out as
simple and efficient techniques for interface reconstruction.
The multigrid algorithm for complex
boundary we propose here can be easily implemented with any of these techniques
as the basic interface reconstruction scheme. 

In what follows, we however simplify the problem even further, since
we work in the framework of a one-way AMR Poisson solver.  At the
finest level (i.e., on the actual grid where the Poisson equation is
to be solved), the boundaries $\partial \Omega$ are simply positioned
at the faces of the outer cells themselves.  In the level set
approach, the boundary could have been placed at the zero level of
some interpolated distance function \citep[see e.g.][]{Gibou2002}.
Since computing distance function comes with a cost, we
have considered here a simpler approach based on a coloring
scheme: inner cells are initialized with a mask function with value
$m_i = 1$ and outer cells with a mask $m_i = -1$. The domain boundary
$\partial \Omega$ is defined at the position where the interpolated
value of the cell-centered mask ($m_i \in [-1, 1]$) crosses zero. At
the finest multigrid level, these interpolated positions are at the
cell faces, as desired.

We now compute the mask value at coarser multigrid levels by simply
averaging recursively the mask value at the finest level (as for the
residual).  Here again, if one wants to use a distance function to
capture the boundary, one could compute the coarser representation of
the distance function also by simple averaging. At each level $\ell$
in the multigrid hierarchy, the domain $\Omega^\ell$ is defined as the
set of cells for which $m^\ell_i > 0$.  The use of the R1 operator
maintains the correct location of the boundary as much as possible,
without spreading the boundary information too much across
neighboring coarse cells.

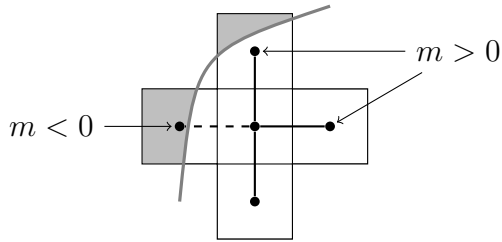
\begin{figure} % {{{ Fig: stencil and wet/dry cells
\begin{center}
\beginpgfgraphicnamed{stencil}
\begin{tikzpicture}

\tikzstyle{legend arrow}=[->, shorten >=1pt]

\newcommand{\stencilpath}{(1,0) -- (1,1) -- (0,1) -- (0,2) -- (1,2) -- (1,3) --
(2,3) -- (2,2) -- (3,2) -- (3,1) -- (2,1) -- (2,0) -- cycle}

\newcommand{\boundpath}{(0.5,0.5) .. controls(0.7, 2.5) .. (2.5,3.1)}
\begin{scope}
\clip \stencilpath;
\draw (0,0) grid (3,3);
\fill[gray!50] \boundpath -- (-5,5) -- cycle;
\end{scope}

\node[fine cell center, black] (wet) at (0.5, 1.5) {};
\node[fine cell center, black] (cen) at (1.5, 1.5) {};
\node[fine cell center, black] (dry1) at (2.5, 1.5) {};
\node[fine cell center, black] (dry3) at (1.5, 0.5) {};
\node[fine cell center, black] (dry2) at (1.5, 2.5) {};

\draw \stencilpath;
\draw[very thick, gray] \boundpath;

\draw[legend arrow] (-0.5, 1.5) node[left] {$m<0$} -- (wet);
\node[right] (mpos) at (3.5, 2.5) {$m>0$};
\draw[legend arrow] (mpos) -- (dry1);
\draw[legend arrow] (mpos) -- (dry2);

\draw[thick] (cen) -- (dry1) (cen) -- (dry2) (cen) -- (dry3);
\draw[dashed, thick] (cen) -- (wet);

\end{tikzpicture}
\endpgfgraphicnamed
\end{center}
\caption{The Laplacian stencil near boundaries: the center of the
  leftmost cell lies outside the domain, and therefore has a negative
  mask value $m<0$. For this cell, a ghost value will be reconstructed
  as shown on Fig. \ref{fig:ghost}.  For all the three other cells,
  the actual cell center values are used unchanged.}
\label{fig:stencil}
\end{figure} % }}}

\begin{figure} % {{{ Fig: Ghost value
\begin{center}
\beginpgfgraphicnamed{ghost}
\begin{tikzpicture}[xscale=1.0,yscale=1.0]
 \foreach \x in {2,4,6} \draw (\x,-0.1)--(\x,0.1);
 \foreach \x in {5,7} \node[fine cell center, black] (\x) at (\x,0) {};
 \foreach \x in {1,3} \node[fine cell center, black, draw, fill=none] (\x) at (\x,0) {};
 \draw (0,0) -- (1) -- (3) -- (5) -- (7) -- (8,0);
 \node[below] at (3,0) {$i-1$};
 \node[below] at (5,0) {$i$};

 % Mask curve
 \tikzstyle{mask}=[black,nodes=fill,circle,inner sep=1pt];
 \path[mask] (1,-1.5) node (m1) {} -- (3,-0.5) node (m2) {} --
 (5,1) node (m3) {} -- (7,1.5) node (m4) {};
 \draw[mask] (m1) -- (m2) -- (m3) -- (m4);
 \node[right] at (m4) {$m$};

 % Boundary position
 \draw[thick,gray] (3.666,-0.3) -- (3.666,0.3);
 \draw[gray,snake=zigzag,segment amplitude=2.5, segment length=3]
 (0.1,0) -- (3.666,0);

 % Real phi
 \tikzstyle{phi} =[blue,nodes=circle,inner sep=1pt];
 \path[phi] (5,-0.5) node[fill] (p3) {} --
 (7,-1.2) node[fill] (p4) {};
 \draw[phi] (p3) -- (p4);
 \node[blue,right] at (p4) {$\Phi$};
 % Ghost phi
 \path[phi] (3,1) node[draw] (p2) {};
 \draw[phi] (p2) -- (p3);
 \node[blue,above] at (p2) {ghost $\tilde{\Phi}_{i-1}$};
 % Boundary phi
 \node[red, fill, inner sep=2pt] (b) at (3.666,0.5) {};
 \node[red, above] at (b) {$\Phi_b$};
\end{tikzpicture}
\endpgfgraphicnamed
\end{center}
\caption{Reconstruction of ghost cell values across boundaries.  Cells
  with negative mask (white dots at centers) are outside of the
  domain and do not carry a valid $\Phi$ value.  The computation of
  the Laplacian at cell $i$ requires the use of a ghost value
  $\tilde{\Phi}_{i-1}$ which is obtained from $\Phi_i$ by linearly
  extrapolating the boundary condition. In accordance with the level
  set idea, the boundary is positioned at the point where $m$ crosses
  zero, which is found by linearly interpolating the mask values.}
\label{fig:ghost}
\end{figure}
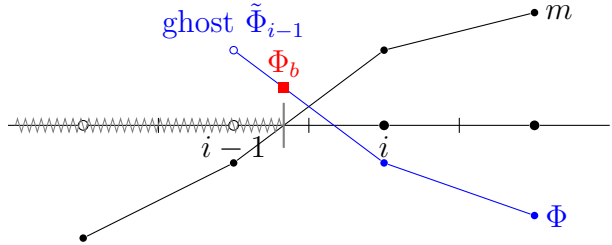

In order to reconstruct the location of the multigrid boundary, and solve for
the corresponding Poisson problem, we use a prescription similar to
the one described in \citet{Gibou2002}. The idea is to redefine the
Laplacian operator close to the domain boundary.  Whenever one of the
neighboring $\Phi_j$ has $m_j \leq 0$ (and therefore lies outside
$\Omega$, see Fig.~\ref{fig:stencil}), it is replaced by a ghost value $\tilde{\Phi}_j$ linearly
extrapolated from $\Phi_i$ and the boundary value.
This ghost value depends explicitly on
$\Phi_i$ and the boundary condition.  A 1D illustration is presented
on Figure~\ref{fig:ghost}.
Since this prescription uses a linear
reconstruction for the multigrid boundary in each direction, the boundary
position is recovered at second order spatial accuracy.
As in \cite{Gibou2002}, the boundary is
positioned in each direction independently, and the 1D case trivially
generalizes to 3D.

The coarsening of the boundary from $\partial \Omega_f$ to $\partial
\Omega_c$ is now fully specified by the restriction operator used for
the mask $m_i$ and by the modified Laplacian closed to the boundary.
These are the 2 key ingredients in our multigrid scheme: the boundary
remains at the same location (up to second order in space) when we go
from fine to coarse levels in the multigrid hierarchy. The boundary
condition $\Phi=0$ is therefore imposed at the correct location, so
that the coarse solution of the Poisson equation with the coarse
boundary corresponds to the solution of the Poisson equation at the
finer level with the fine representation of the boundary.

We speculate that the chosen multigrid boundary reconstruction scheme should
satisfy an order condition similar to the prolongation and restriction
operators \citep{Wesseling1992, Press1992}. Although a mathematical proof is
beyond the scope of this paper, our numerical experiments suggest that
this is indeed the case.  We see in Figure~\ref{fig:rate} that, in
case of a smooth boundary, our second-order multigrid boundary reconstruction
scheme results in a perfect multigrid convergence\footnote{Multigrid
  convergence means here that the damping rate of the residual norm
  does not depend on the grid size.}, while the first-order scheme
doesn't. As we will now discuss, in case of very complex boundaries,
second-order reconstruction of the multigrid boundary is not possible anymore,
and we have to degrade our scheme to first-order to ensure
convergence.

 % }}}

\section{First-order multigrid boundary reconstruction and the ``small islands'' problem}
\label{sec:Islands}

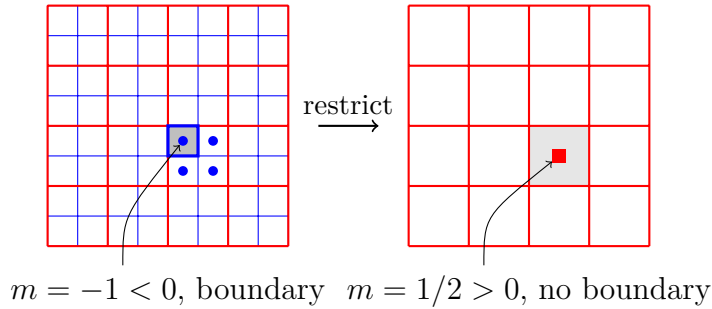
\begin{figure} % {{{ Fig: Small island
\begin{center}
\beginpgfgraphicnamed{island}
\begin{tikzpicture}[scale=0.4]
  \begin{scope}
	\draw[fine cells] (0,0) grid (8,8);
	\draw[coarse cells] (0,0) grid[step=2] (8,8);

	\draw[fine boundary, very thick, fill=gray!50] (4,3) rectangle (5,4);
	\node[fine cell center] (f) at (4.5, 3.5) {};
	\node[fine cell center] at (5.5, 3.5) {};
	\node[fine cell center] at (4.5, 2.5) {};
	\node[fine cell center] at (5.5, 2.5) {};

	\node (label1) at (2.5, -1) {};
	\node at (4.0, -1.5) {$m=-1<0$, boundary};
	\draw[->] (label1) .. controls (2.5, 1) .. (f);

	\draw[thick, ->] (9, 4) -- node[pos=0.5, above] {\small restrict} (11, 4);
  \end{scope}

  \begin{scope}[xshift=12cm]
	\fill[gray!20] (4,2) rectangle (6,4);
	\draw[coarse cells] (0,0) grid[step=2] (8,8);
	\node[coarse cell center] (c) at (5, 3) {};

	\node (label2) at (2.5, -1) {};
	\node at (4.0, -1.5) {$m=1/2>0$, no boundary};
	\draw[->] (label2) .. controls (2.5, 1) .. (c);
  \end{scope}

\end{tikzpicture}
\endpgfgraphicnamed
\caption{Coarsening (restriction) may cause loss of boundary
  conditions. The Dirichlet BCs around an isolated cell with $m=-1$
  among a sea of $m=1$ cells are lost after one coarsening step, as
  the resulting coarse mask is positive everywhere.}
\label{fig:island}
\end{center}
\end{figure}

\begin{figure*}
  \begin{center}
 \begin{tabular}{cc}
 \includegraphics[width=0.45\linewidth]{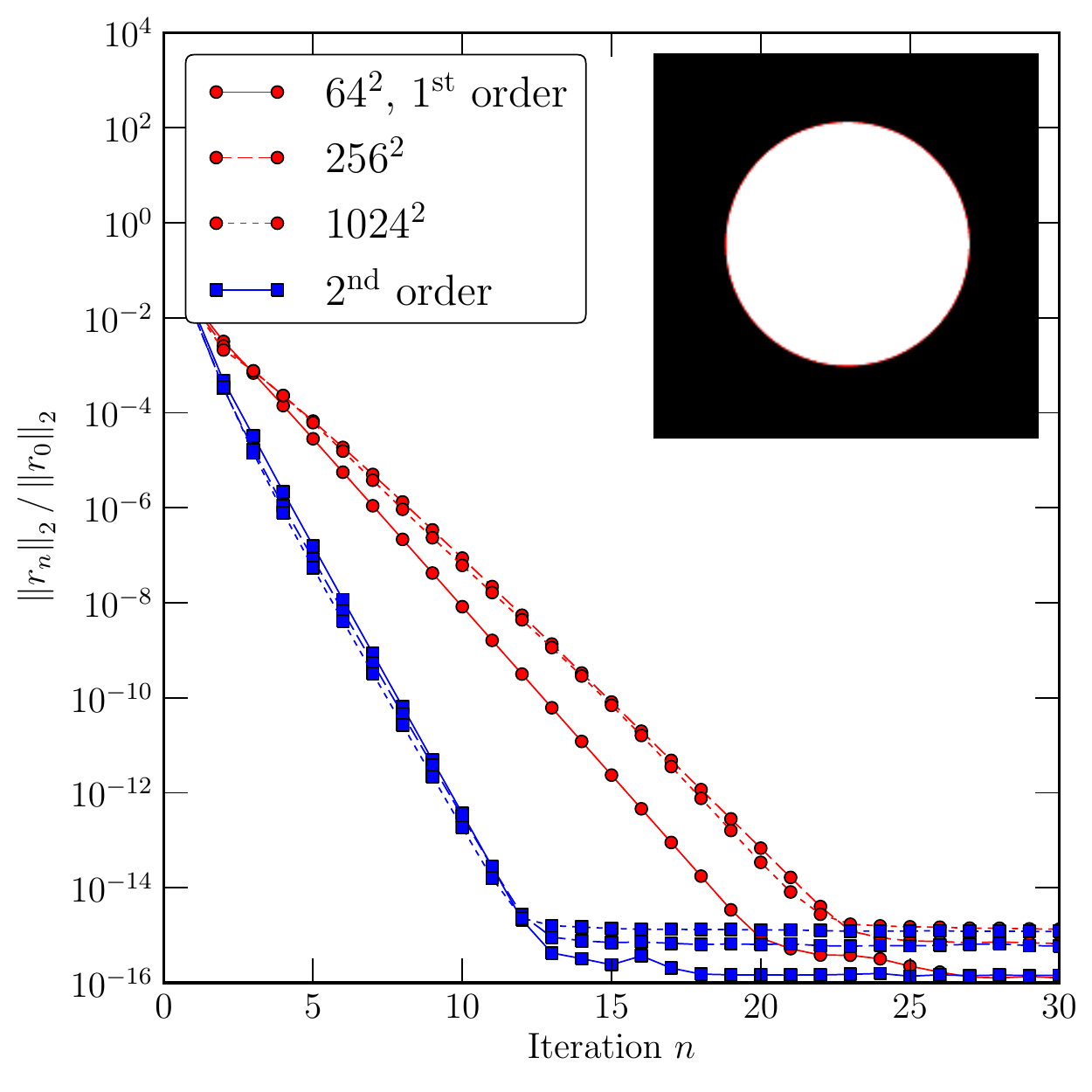} &
 \includegraphics[width=0.45\linewidth]{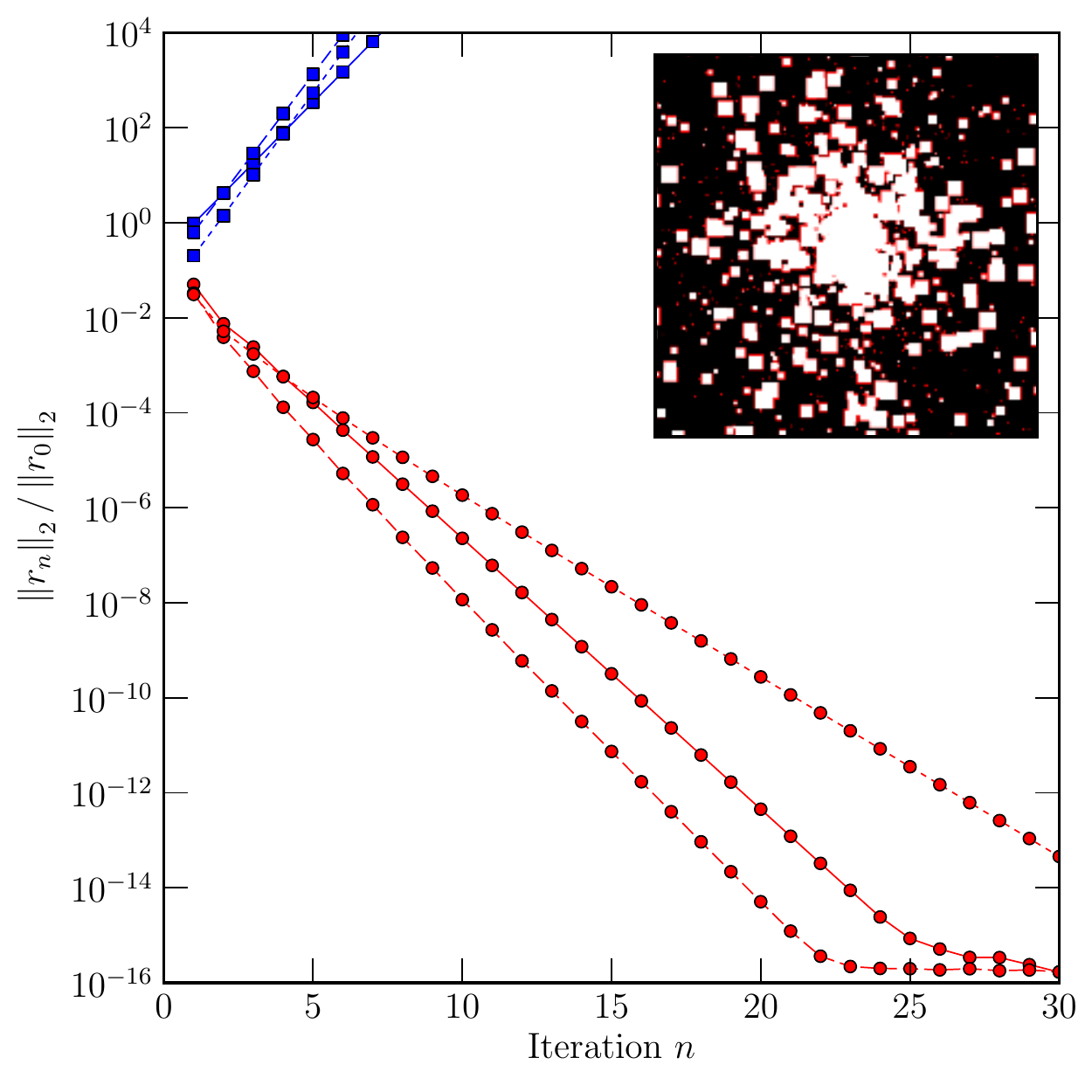}
 \end{tabular}
 \end{center}
 \caption{$L_2$ norm of the residual as a function of iteration number
   in our multigrid Poisson solver in 2D for both presented boundary
   capturing schemes. For each domain shape shown in the insets, the residual for
   both 2\textsuperscript{nd} order (in blue) and 1\textsuperscript{st} order
   (in red) boundary capturing schemes is presented for a $64^2$ grid 
   up to a $1024^2$ grid.}
 \label{fig:rate}
\end{figure*}

We see in Figure~\ref{fig:rate} that a different type of boundary can
lead to a catastrophic divergence of our multigrid scheme (the blue
line shows the residual norm evolution in case of second-order
multigrid boundary reconstruction). This rather complex boundary condition is
typical of AMR grids in cosmological simulations
\citep{Teyssier2002}. It features many small disconnected domains that
cluster in a large central region with many ``holes''.  Successive
coarsening of the grid resolution leads in this case to a loss of
boundary conditions, especially when ``holes'' or ``small islands''
are present.  Indeed, in this case, the finer boundary small scale
features cannot be represented anymore on coarser
grids. Figure~\ref{fig:island} illustrates such a case, for which the
fine grid still has a cell with a negative mask value, so that the $\Phi=0$ boundary
condition still applies, but the coarse grid has cells with only positive values.
Because of this topological change in the boundary representation, the
coarse and fine solutions of the Poisson equation become significantly
different.  Spurious eigenmodes associated to this loss of constraints
at the boundary are not damped quickly enough by the smoothing operator at the fine
level: this leads to slower convergence rates and, in some case, to
catastrophic divergence (see Fig.~\ref{fig:rate}).

 % }}}

One previously proposed solution to this problem is the subtraction of
the divergent modes introduced by these small islands, for example by
recombining the multigrid iterants \citep{Brandt1995}. This however
turns out to be too computationally and memory intensive for us to use
in the Poisson solver of the RAMSES code: it requires storing a large
number of previous solutions.  This recombination method also tends to
perform worse when the number of islands increases, which would likely
render it useless in the case of the complex grid structure
encountered in cosmological simulations.

Another proposed solution is to stop coarsening the residual in the
multigrid hierarchy whenever such a problem occurs during multigrid coarsening
\citep{Johansen1998}.  This degrades the performance of the multigrid
method, since large scale modes are no longer damped efficiently.  In
the case of very complex boundaries with small islands, like in
Figure~\ref{fig:rate}, the level at which the hierarchy has to be
truncated is so close to the finest level that the whole multigrid
approach breaks down.

In order to overcome these limitations, \citet{McCorquodale2001}
proposed to keep track of the boundary representation at the fine
level, and modify the Laplacian on coarse grids, whenever the
operator stencil crossed this fine boundary.
Stencil nodes which cannot be reached from the stencil center by a 
straight segment without crossing the boundary are excluded, and an asymmetric stencil is used, chosen in function of the local configuration of the interface.

We propose here a simpler approach, based on the observation that
these small islands correspond to local minima in the color function
$m_i$ (or equivalently in the distance function if needed). The
averaging scheme will tend to smooth these extrema, resulting in the
disappearance of the negative values (see Fig.~\ref{fig:island}) and in
the apparition of spurious boundary conditions. In analogy to
traditional high-order numerical schemes for hyperbolic systems of
conservation laws \citep{VanLeer1984,Colella1984}, we solve
the problem by switching {\it globally} to a first-order multigrid boundary
reconstruction scheme.  We impose the $\Phi=0$ constraint at the
coarse level on the grid point nearest\footnote{The so-called NGP
  scheme for Nearest Grid Point.} to the interface: in practice, for
each cell for which the mask value $m_i < 1$, we impose $\Phi_i=0$.
The simple averaging restriction for the mask ensures that cell $i$
has $m_i < 1$ if and only if it has a non-zero intersection with the
boundary.

We have tested our first-order multigrid boundary reconstruction scheme on 2
extreme types of boundary conditions~: a simple disk with no hole or
island, and a complex clustered grid, typical of AMR cosmological
simulations. We have compared its convergence properties to the
second-order reconstruction scheme in Figure~\ref{fig:rate}.  In the
disk case, the simple topology of the boundary allows us to take full
advantage of the second-order reconstruction of multigrid boundaries,
resulting in fast convergence rates (in blue), totally independent of
the problem size. This shows that our overall scheme features a true
``textbook'' multigrid convergence.  Using the first-order multigrid boundary
reconstruction, we degrade the convergence rate significantly, and it
now depends slightly on the problem size. This is expected since the
position of the boundary at two consecutive multigrid levels can differ
by half a cell width, and this first-order error in the boundary
positioning limits the accuracy of the coarse correction.

In the complex boundary case, with many holes and islands, our
second-order reconstruction fails, as we can see from the
divergent behavior of the residual norm in Figure~\ref{fig:rate}. We
interpret this catastrophic behavior as follows: topological changes
in the boundary representation at coarse levels result in spurious
long-range mode that are not damped by the smoothing operators at
finer levels. If we use first-order multigrid boundary reconstruction, we
observe that the convergence is restored: although first-order reconstruction 
of the boundary introduces
small-scale errors close to the boundary, these errors are efficiently
damped by the smoothing operator at finer levels. 
Note however that the convergence rate is slower than 
in the disc case, and that the convergence rate now depends on the grid
size. A similar effect was observed in \cite{Day1998, Popinet2003}.

Note that in our present implementation, we have to decide beforehand
to which order we wish to reconstruct the multigrid boundaries.  We have tried to
implement an adaptive algorithm, for which the reconstruction scheme
is adapted locally to the topological changes of the boundary, but no
satisfactory results were found.  This then raises the question: how do
we choose between first or second order reconstruction, when solving
the Poisson equation on an arbitrarily complex grid? It may be
possible to decide from a topological analysis of the grid (such as
its genus and the number of connected components), but this proves
difficult to achieve in practice.

We have opted for a pragmatic approach, and decide at run time 
which scheme to employ for each AMR level independently.  When we
start solving the Poisson equation on a given level, we first try
using the second-order reconstruction. We monitor the convergence rate
during the iterations, and if it becomes slower than a fixed
threshold, i.e. if $\left\| r_{n+1} \right\|_2 / \left\| r_n \right\|_2 >
\eta$ with typically $\eta = 0.5$, we switch to the first-order scheme
for that level only and for the next 10 solves.  With our current
implementation, if the AMR grid is really complex (with small
islands), the solver only wastes a couple of iterations before deciding on which of
first or second-order gives the best convergence rate.  This
works very well in practice, even in cosmological simulations
featuring clusters and filaments, as only a few intermediate AMR
levels exhibit very complex topologies.

\section{Accuracy and performance tests}

\subsection{Accuracy tests}

We have tested the accuracy of the solver by comparing numerical solutions computed using RAMSES to exact analytical expressions.
% The setup
We chose a 2D setup inspired from galaxy simulations, with a radial mass distribution centered in the computational box.
With coordinates $(x,y) \in [0,1]^2$ the radial coordinate is given by $r^2 = (x-\frac{1}{2})^2 + (y-\frac{1}{2})^2$.
We take the analytic potential $\Phi_e(r)$---subscripted with $e$ for exact---to be
\begin{equation}
	\Phi_e(r) := \ln \left( \left[\frac{r}{r_0}\right]^2 + 1 \right),
	\label{eq:phi-trunc}
\end{equation}
which is smooth everywhere in the domain and features a core at the origin.
The parameter $r_0$ controls the concentration of the profile.
The corresponding density profile $\rho_e(r)$ is obtained from \eqref{eq:poisson}:
\begin{equation}
	\rho_e(r) = \frac{4 r_0^2}{\left(r^2 + r_0^2\right)^2},
	\label{eq:rho-trunc}
\end{equation}
which is positive and smooth everywhere.
The analytical expression for the force intensity $f_e = |\nabla \Phi|$ is
\begin{equation}
	f_e(r) = \frac{2r}{r^2 + r_0^2},
	\label{eq:force-trunc}
\end{equation}
and the force vector everywhere points towards the center of the box.
The boundary conditions at the border of the computational box are set using the exact solution \eqref{eq:phi-trunc}.

Using this setup, we have computed the truncation error for the potential and the force as a function of the finest grid resolution for both a Cartesian grid and an AMR grid with 3 additional AMR levels.
For the resolution level $\ell$, the finest cell size $\Delta x$ in the grid is given by $\Delta x = (1/2)^\ell$.
We use a quasi-Lagrangian mesh refinement strategy, closest to what is commonly employed in astrophysical applications:
given a fixed threshold mass $M$, each cell containing a mass exceeding $2^{N_\mathrm{dim}} M$ is refined (split) into $2^{N_\mathrm{dim}}$ children cells.

\begin{figure}
	\centering
	\includegraphics[width=0.45\linewidth]{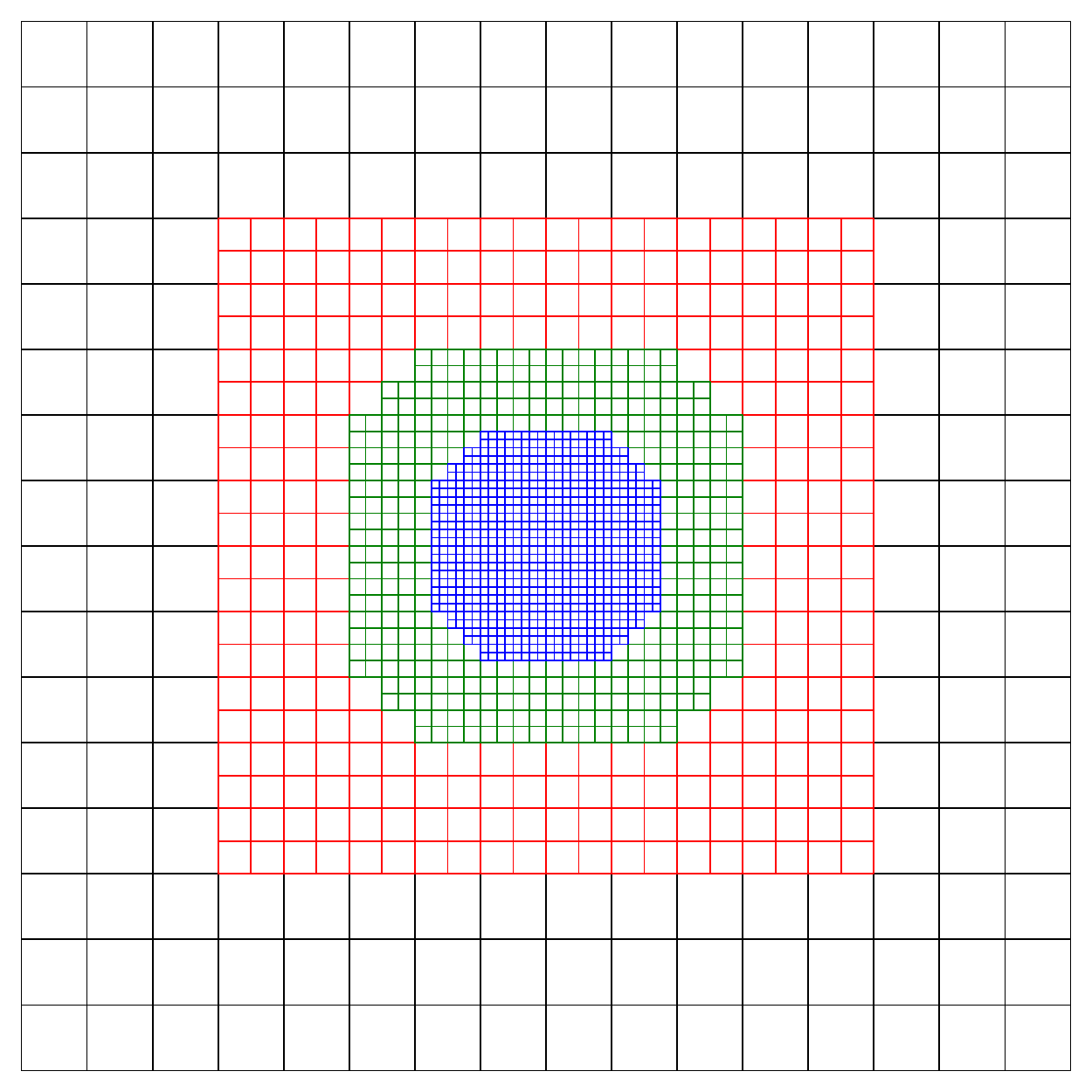}
	\caption{AMR grid in the accuracy test case described in the text, at resolution level 7.
		The base grid is at level 4, with 3 AMR levels reaching down to level 7 using a quasi-Lagrangian refinement strategy.}
	\label{fig:trunc-grid}
\end{figure}

In our test, whenever the resolution is increased by 1, the base (coarsest) AMR level is incremented by 1, and the mass resolution $M$ is divided by 4 (in 2D). This procedure allows us to increase the resolution of the finest AMR cells by exactly one level, while keeping the depth of the AMR grid a constant.
We initially pick the value of $M$ such that it triggers the refinement of 3 AMR levels.
The resulting grid for resolution level $7$ is shown on Fig.~\ref{fig:trunc-grid}.
\begin{figure*}
 \begin{center}
 \begin{tabular}{cc}
 \includegraphics[width=0.45\linewidth]{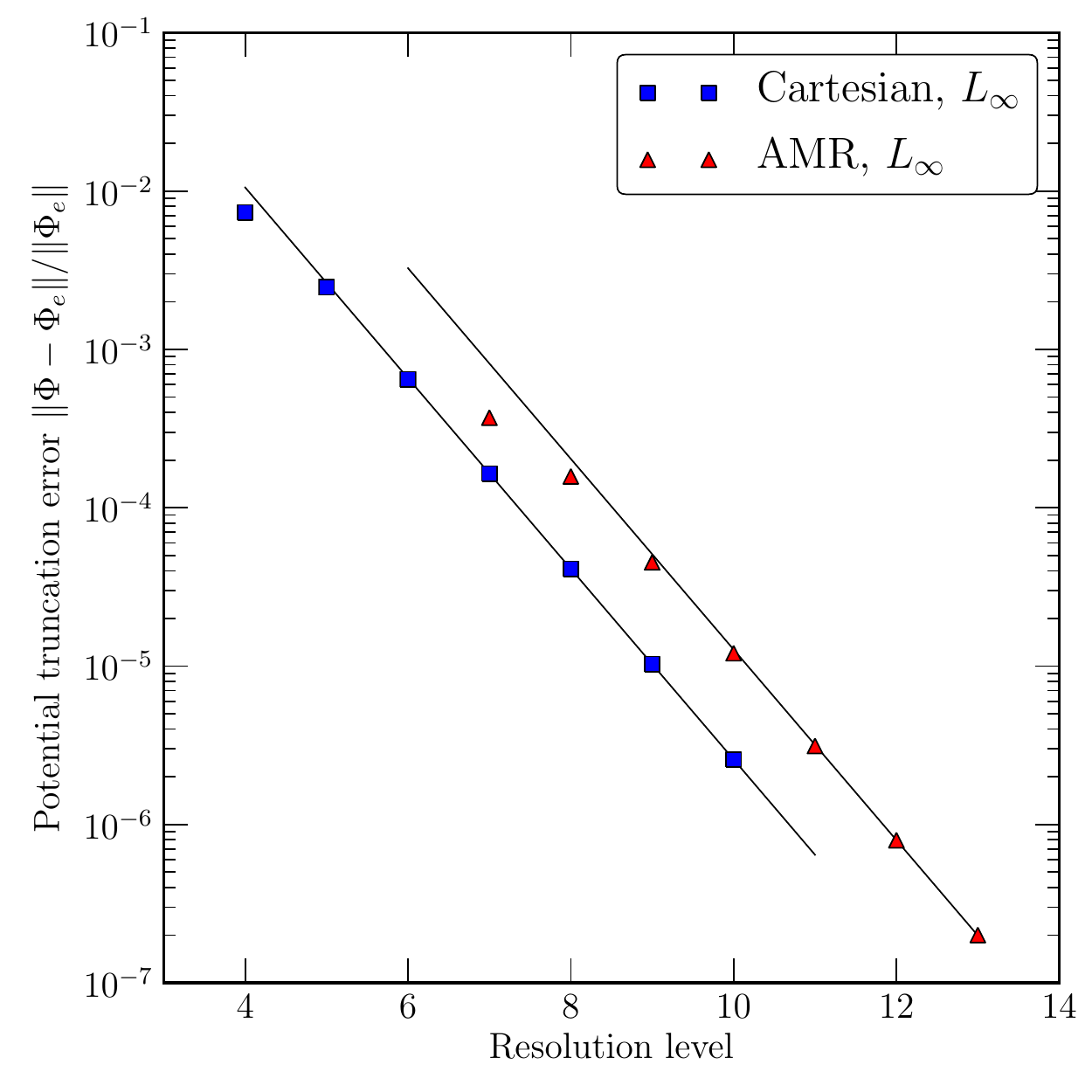} &
 \includegraphics[width=0.45\linewidth]{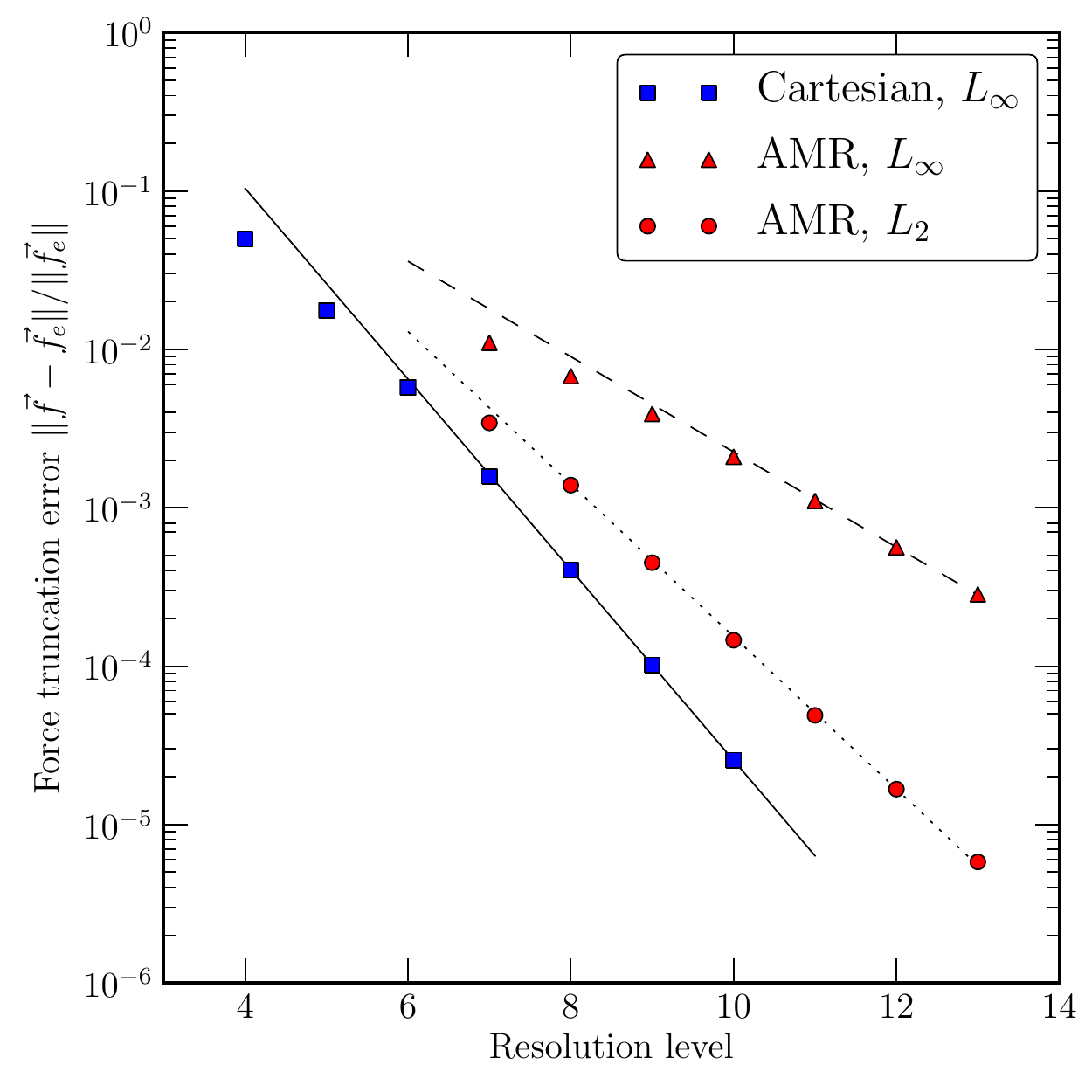}
 \end{tabular}
 \end{center}
 \caption{Truncation error on the potential (left) and force (right) for the setup described in the text,
		as a function of the maximal resolution level $\ell$. The finest cell size $\Delta x$ is given by $\Delta x = (1/2)^\ell$.
		The solid lines are $\mathcal{O}(\Delta x^2)$ (second order), the dashed line is $\mathcal{O}(\Delta x)$ (first order),
		and the dotted line is $\mathcal{O}(\Delta x^{1.6})$.
	}
 \label{fig:trunc}
\end{figure*}

We have evaluated the truncation errors for both the potential and the gravitational force in order to test the one-way interface scheme.
Once a solution for the potential has been obtained at for a given AMR level, RAMSES computes the force for this level using a 5-point finite difference approximation of the gradient:
\begin{equation}
	\partial_x \Phi = \frac{4}{3} \frac{\Phi_{i+1}-\Phi_{i-1}}{2 \Delta x}
			- \frac{1}{3} \frac{\Phi_{i+2} - \Phi_{i-2}}{4 \Delta x} + \mathcal{O}((\Delta x)^4) 
	\label{eq:force-fda}
\end{equation}
\begin{figure}
	\centering
	\beginpgfgraphicnamed{force-fda}
	\begin{tikzpicture}[scale=1.1]
		\draw[fine cells] (0,0) grid (4,4);
		\draw[coarse cells, dotted, thin] (5, 0) -- (5, 4);
		\draw[coarse cells, dotted, thin] (4, 3) -- (6, 3);
		\draw[coarse cells, dotted, thin] (4, 1) -- (6, 1);
		\draw[coarse cells, scale=2] (0,0) grid (3,2);
		\foreach \y in {0.5, 1.5, 2.5, 3.5} {
			\node[boundary point] at (4, \y) {};
		}
		\draw[fine cells, very thick] (1.5,1.5)--(5.5,1.5);
		\foreach \x in {1.5, 2.5} {
			\node[fine cell center] at (\x, 1.5) {};
		}
		\foreach \x in {4.5, 5.5} {
			\node[fine cell center, draw, fill=white, thick] at (\x, 1.5) {};
		}
		\node[fine cell center, inner sep=2pt] at (3.5, 1.5) {};
		\node[above] at (3.5, 1.5) {$\partial_x \Phi$};
	\end{tikzpicture}
	\endpgfgraphicnamed
	\caption{Computation of the gravitational force in RAMSES using the finite difference approximation \eqref{eq:force-fda}.
		At a fine--coarse interface, the missing fine values (empty circles)
		are computed from the coarse level potential by linear interpolation.
		Prior to solving for $\Phi$ at the fine level, the boundary values for the fine Poisson problem
		(blue crosses, see also Fig.~\ref{fig:grid}) had been computed in the same manner.
	}
	\label{fig:force-ramses}
\end{figure}
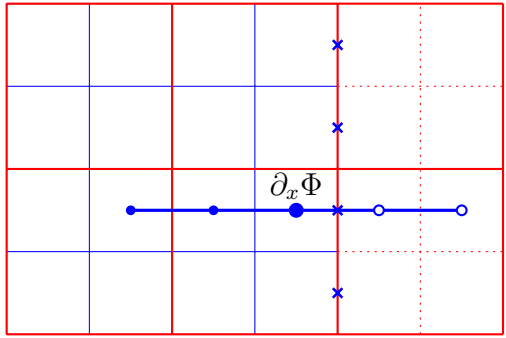

The truncation errors are presented on Fig.~\ref{fig:trunc}.
For the potential, both Cartesian and AMR grids feature second-order convergence.
For a given resolution level, the AMR truncation error is larger than the corresponding Cartesian error:
this is expected, since AMR will only use the maximal resolution in particular areas of the grid, leading to bigger truncation errors because of the presence of coarser cells.

Similarly to the potential, in the Cartesian case, the truncation error of the force decreases at second order.
In the AMR case however, the convergence is degraded to first order for the $L_\infty$ norm.
We attribute this effect to the one-way interface scheme.
Whenever the finite difference approximation of the gradient crosses the level boundary,
RAMSES fills in the missing values by linear interpolation from the coarser level, as illustrated on Fig.~\ref{fig:force-ramses}.
Since the coarse Laplacian operator and linear interpolation are both accurate to second order,
the interpolated values (empty circles on Fig.~\ref{fig:force-ramses}) have a truncation error of the form $(\Delta x)^2 \epsilon_i$, where $\epsilon_i=\mathcal{O}(1)$.
Note that $\epsilon_i$ accounts for both the coarse Laplacian and interpolation errors.
The valid values at the fine level (solid dots on Fig.~\ref{fig:force-ramses}) are accurate to second order,
with a truncation error $(\Delta x)^2 \eta_i$, with $\eta_i=\mathcal{O}(1)$.
Across the interface, the truncation error jumps abruptly from $(\Delta x)^2 \epsilon_i$ to $(\Delta x)^2 \eta_i$.
Since there is no reason for $\epsilon_i$ and $\eta_i$ to connect to the same value at the coarse--fine boundary,
the difference formula \eqref{eq:force-fda} will produce a $\mathcal{O}(1/\Delta x)$ spike on the derivative of these terms.
This translates to a $\mathcal{O}(1/\Delta x) (\Delta x)^2 = \mathcal{O}(\Delta x)$ error on the gradient of $\Phi$.

This problem was previously noted in the case of AMR solvers \citep[see e.g.][for a discussion for an AMR Poisson solver]{Martin1996}, and is inherent to the one-way interface scheme.
It could be avoided using higher order Laplacian and interpolation operators.
Using third order operators would reduce the error at the interface to $\mathcal{O}{(\Delta x)^2}$.

\begin{figure*}
	\centering
	\includegraphics[width=0.45\textwidth]{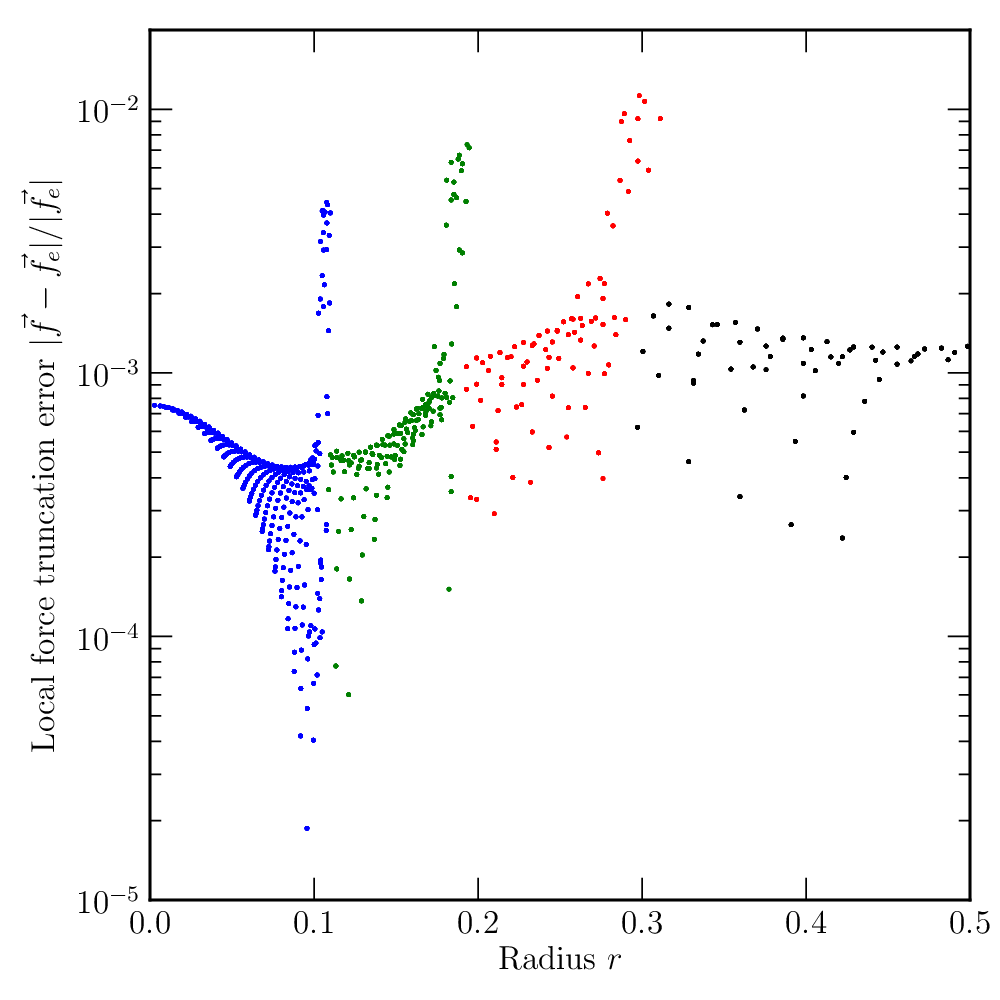} \hfill
	\includegraphics[width=0.45\textwidth]{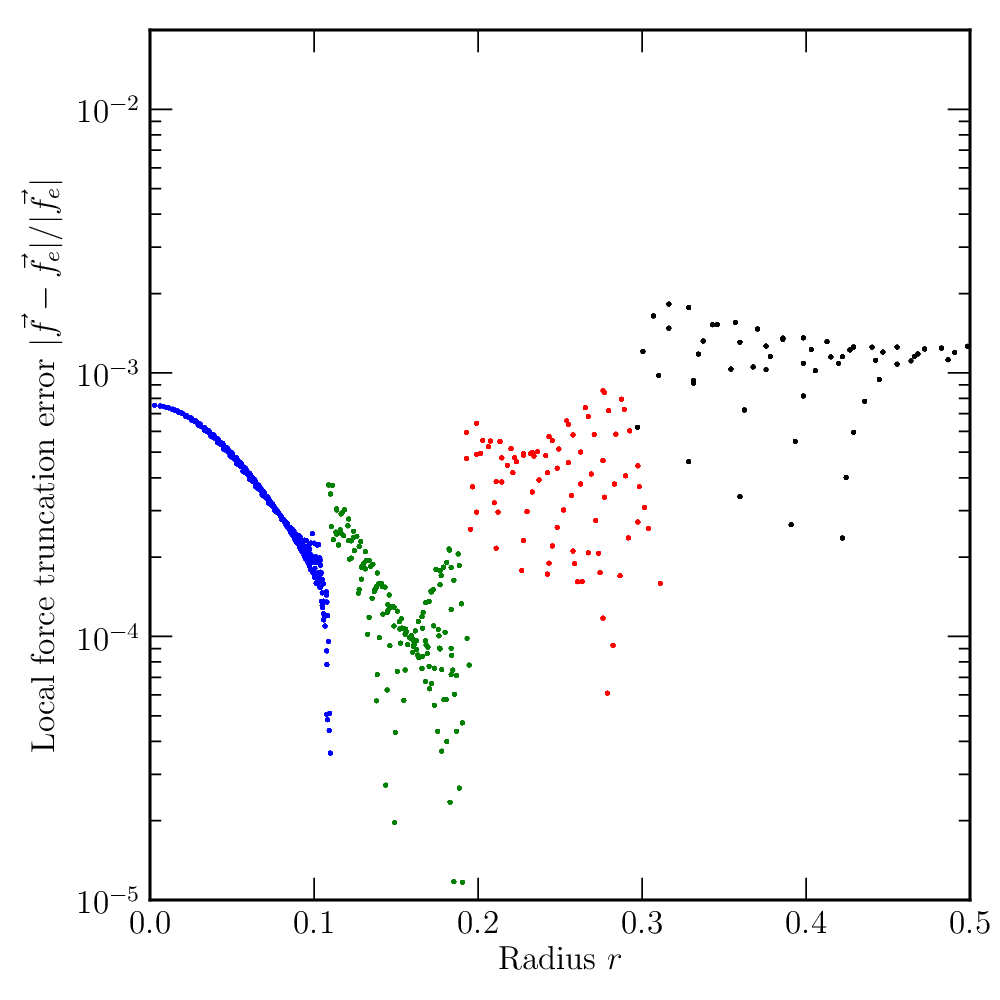}
	\caption{Relative error on the force in the 3-level setup described in the text, at resolution level 8.
		The values at cell centers are colored by level correspondingly to Fig.~\ref{fig:trunc-grid}.
		On the left panel, the border potential at fine--coarse boundaries is interpolated from the coarser level, as implemented in RAMSES with the one-way interface scheme.
		The interpolation results in a degradation of the force to order 1 within shallow regions around the AMR level transitions.
		On the right panel, the border potential is obtained from the analytic solution, illustrating a best-case solution on the given grid.
	}
	\label{fig:trunc-force-radial}
\end{figure*}

Fig.~\ref{fig:trunc-force-radial} represents the local force error as a function of cell radius for the test problem at resolution level 8.
The left panel shows the RAMSES prescription using interpolation for computing potential and force at level boundaries, according to Fig.~\ref{fig:force-ramses}.
In the right panel, the missing potential values for both boundary conditions and force computation are no longer interpolated from the coarse level,
but rather evaluated directly from the exact analytic solution, which suppresses any truncation error problem at the level interfaces.
The comparison of the two panels shows that the impact of the one-way interface scheme on the global quality of the solution is minimal:
The local force error never exceeds about 1\%, and the first order degradation of the force is confined to thin shells of codimension 1.
As a result, the $L_2$ convergence of the force is still close to order 2, scaling approximately as $\Delta x^{1.6}$ (see Fig.~\ref{fig:trunc}).

Note that although this test case is simple, it provides valuable insight on the properties of the one-way interface scheme
coupled to the computation of the force.
In the case of RAMSES, extensive tests of force in the code had been made for cosmological simulations in \citet{Teyssier2002}
using the original Poisson solver.
Errors on the force due to the one-way interface AMR were typically found to be within 1\%.
Since the Laplacian operator is identical \emph{at the fine (AMR) level} in our multigrid implementation to the one used for these tests,
our solver would yield exactly the same solution when fully converged to numerical accuracy.

\begin{figure}
	\centering
	\includegraphics[width=0.45\textwidth]{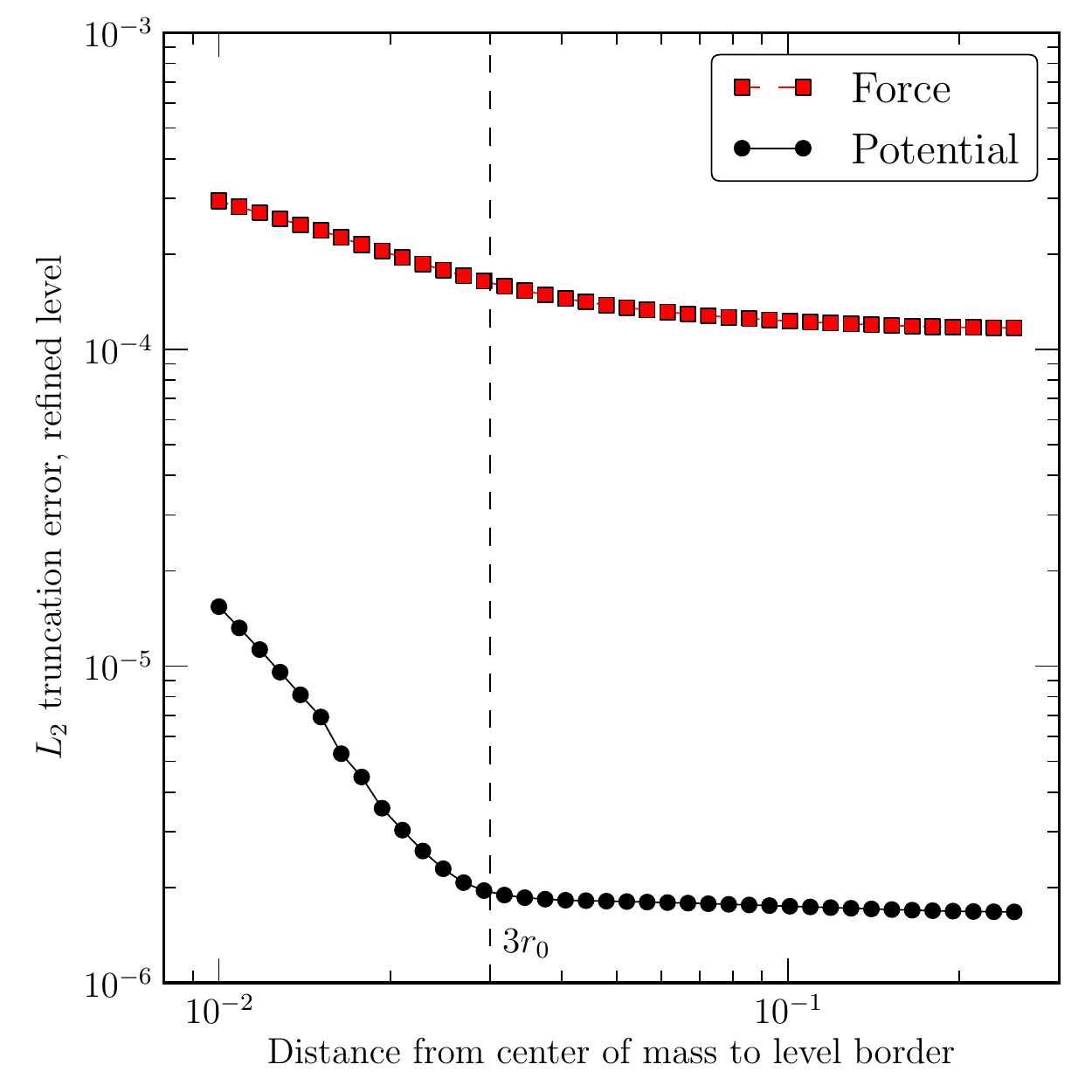}
	\caption{Effect of the gradient at a level boundary on the quality of the fine level solution,
		for the potential (circles) and the force (squares).
		For boundary--center of mass separations greater than about $3r_0$, the location of the density peak has no sizable effect.
		As the peak gets closer to the boundary, stronger gradients introduce errors in the numerical potential.
		Note however that the errors on the force---which are dominated by the first-order terms arising at the level boundary---%
		are comparatively insensitive to the boundary gradient.
		}
	\label{fig:grad-error}
\end{figure}

Finally, we have tested the convergence and accuracy of the solver in the presence of gradients at a level boundary.
We used a concentrated mass distribution, given by Eq.~\eqref{eq:rho-trunc} with $r_0=0.01$, centered in the computational box.
Starting from a base resolution at level 10 (corresponding to a regular grid of $1024^2$ cells), we introduced one level of refinement at level 11, delimited by a circular boundary of radius 0.25.
By moving around the refined region within the box, we can probe the effect of the gradient on the solver as the central peak in the mass distribution approaches the boundary.
For various positions of the refined level, we have performed multigrid iterations with our algorithm with a residual reduction factor set to $10^{-7}$.
The resulting truncation error for the fine level is plotted on Fig.~\ref{fig:grad-error} as a function of the boundary--mass center distance.

At separations greater than about $3r_0$, the gradient induced by the mass distribution has no noticeable effect on the quality of the global solution, for both the potential and the force.
As the mass distribution gets closer to the level boundary however, large derivatives at the interface will introduce errors in the Dirichlet boundary condition,
which will impact the quality of the solution in the whole refined region.
Note, nevertheless, that in our test the force seems much less sensitive to this effect,
possibly because its truncation error is already dominated by the first order term arising at level boundaries.

While this effect is inherently present in one-way interface methods, it could likely be reduced by using higher-order reconstructions of the Dirichlet boundary condition.
As evidenced by the results on the force, however, it is not clear that much can be gained by such a sophistication, unless the order of the whole scheme is increased globally.
Rather, these results highlight the importance of selecting suitable AMR refinement criteria.
In computational astrophysics, AMR grids are often refined whenever the local density crosses a threshold.
RAMSES combines such criteria with a level dilation strategy---which ``grows'' AMR level outwards,
and helps reducing the granularity of the grid while limiting sharp variations of the mass distribution along the boundary.

In terms of convergence rate, the boundary gradient had no noticeable effect, as the solver converged in 6 or 7 iterations for all of the data points represented on Fig.~\ref{fig:grad-error}, yielding a convergence factor of about 10.

\subsection{Performance in cosmological simulations}

The new Poisson solver has by now been used in RAMSES in a variety of
simulations.  We present timings for our new multigrid
(MG) solver on Table~\ref{tab:timings}, together with
corresponding timings for the conjugate gradient (CG)
method, used here as a reference example of traditional iterative solvers.
The timings represent the total wall clock time required to solve
for the Poisson equation on the whole AMR grid, for a residual $L_2$ norm reduction factor of $10^{-3}$.
For the reference timings, we have set the initial guess of the solution
to zero everywhere on the grid, for both solvers.

The tests were run on the CEA/CCRT Platine computer, consisting of BULL 
NovaScale 3045 units totaling 932 nodes, networked by an InfiniBand interconnect.
Each node hosts 4 Intel\textsuperscript{\textregistered} Itanium\textsuperscript{\textregistered}~2 
dual-core 1.6~GHz processors sharing 24~Gb of RAM.

\begin{table*}
\begin{center}
\begin{tabular}{llllll}
\hline
\textbf{Type of AMR grid} & \textbf{PE} & \textbf{Resolution} & \textbf{$N_{\mathrm{cells}}$} &
\textbf{CG (s)} & \textbf{MG (s)} \\ \hline
A. Cartesian & 32 & $256^3$ + 0 levels & 19M & 52 & 4.1 \\
B. Zoom-in & 64 & $128^3$ + 3 levels & 47M & 160 & 15 \\
C. Cosmology & 1024 & $1024^3$ + 5 levels & 4.8G & 2750 & 1070 \\ \hline
\end{tabular}
\caption{Poisson solver timings on three reference simulations, for
  conjugate gradient (CG) and multigrid (MG).}
\label{tab:timings}
\end{center}
\end{table*}

Simulation A (a cosmological simulation at very early time) has a
base $256^3$ grid, with no additional level of refinement, on 32 processors.
Simulation B is a ``zoom'' simulation, with a base resolution of
$128^3$ and successive forced refinements up to a $1024^3$-equivalent
zoom-in area.  The
cosmological computation follows the formation of a dense dark matter
halo within the focus area.
Simulation C is a clustered, $50 h^{-1} \mathrm{Mpc}$ box cosmological
simulation at $z=1$, with $1024^3$ particles. At this level of
nonlinearity, the finer AMR levels are extremely clustered, while the
intermediate levels exhibit a complex topology as they follow the
intermediate-density structures (walls, filaments and clumps).

All timings of Table~\ref{tab:timings} show a strong performance advantage of the
multigrid method over the conjugate gradient.  Additionally, during the time
evolution of cosmological simulations, we witnessed that the multigrid solver
convergence times are much more predictable and consistent across different runs than the
conjugate gradient.  We attribute this effect to the fact that the multigrid
performance only depends on the topology of the grid, which changes
progressively during the simulations, whereas the conjugate gradient is very
sensitive to the quality of the first guess.

In the context of one-way interface solvers on AMR grids, we can significantly improve 
the performance of the conjugate gradient solver by computing a first guess solution based on
 the coarser level solution.  We choose to
linearly interpolate the initial guess of $\Phi$ at level $\ell$ from the
solution at the coarser level $\ell-1$, which has just been computed.
This ``multilevel'' approach guarantees an initial guess of reasonable quality at a small
extra cost---the cost of interpolating the solution from the coarser AMR level to the finer level.
Note that for our CG implementation, iterations only take place at the
finest level, and are therefore \emph{not} multigrid-accelerated.
We now discuss timings for our 3 test simulations using this improved CG solver. 
Since new multigrid timings have shown to be practically unchanged down to
2~digits, when using this new first guess, we only give new timings for the
conjugate gradient solver.

Simulation A features a CG time between 5.8 and 23~seconds. The conjugate
gradient timings are particularly erratic on the first few timesteps, because
the first cosmological structures form at very small scale in the middle of a
nearly uniform density field; therefore such small scale features are not
accurately represented on the first guess obtained by interpolating the coarse
solution. The number of iterations necessary to reach a given residual
reduction factor is therefore high at the start of the simulation, before
decreasing significantly as larger structures grow.  In any case, the multigrid method
performs significantly better than the conjugate gradient on cartesian grids,
even though the conjugate gradient benefits dramatically from an optimal first
guess, and has less overall overhead.

On simulation B, the CG solver with the new first guess takes 140~seconds.
The almost tenfold performance gain of the multigrid algorithm over the
conjugate gradient can be explained by the fast evolution of the matter density
at the coarse levels in the early stages of the simulation.  Since coarse
levels use a bigger time step than finer levels because of adaptive
timestepping, the potential on coarse cells---which is interpolated as a first
guess for the finer potentials---is updated less frequently.  The finer first
guesses thus tend to be out of synchronization with the real solution,
resulting in additional conjugate gradient iterations.  The multigrid algorithm
is much less sensitive to first guess quality, resulting in a significant
advantage over the conjugate gradient. This situation shows the real strength
of the new solver in the context of adaptive time stepping.

Finally, in the case of simulation C, the CG solver runs for 850~seconds, about
20\% less than multigrid.  Decomposing the solver time by level shows that the MG
solver spends most of its time dealing with very fine and very clustered grids,
at the finest end of the AMR structure.  This is easily understood, as this
type of grid geometries represent a worst-case scenario for the multigrid
solver in terms of small islands, forcing intermediate AMR levels into the
slightly less efficient 1\textsuperscript{st} order capturing mode.  Moreover,
at this stage of the simulation, the timestep is usually extremely small, which
is beneficial to the conjugate gradient as discussed in the case of simulation
B.  This result suggests using a hybrid scheme in practice, where one uses the
new multigrid method for most levels of the AMR hierarchy except the finest
ones, which can be handled by the CG solver.  This method has been implemented in
RAMSES.

For most astrophysical applications, it is sufficient to improve the solution on each level
until the residual norm reaches about $10^{-3}$ times the norm of the initial solution, which is obtained by interpolation of the coarse solution.
Note that we use $N_{pre}=N_{post}=2$.
In the vast majority of practical 3D grid geometries encountered in cosmology and galaxy formation simulations,
the multigrid algorithm performs 4 to 5 iterations, corresponding to convergence factors of about 4--6.
We observe this behavior regardless of the resolution of the simulation, from simple $128^3$ runs to full-scale simulations
starting at $1024^3$ with deep AMR grids.
In this sense, the multigrid performance is close to textbook multigrid convergence in practical situations.

\subsection{Effect of $N_{cycles}$}
\label{sec:cycles}

The $N_{cycles}$ parameter controls how many multigrid iterations are
performed when computing a coarse correction, at any level of the MG
hierarchy. More iterations usually yield a better correction (and less
multigrid iterations at the finest level before reaching tolerance),
but significantly increase the cost of each iteration.  One must
therefore find an appropriate trade-off.

Performing more than 2 or 3 cycles is usually not desirable, because
the coarse problem is only itself an approximation of the fine
correction problem.  We have studied $N_{cycles}=1$ (``V-cycles''),
$N_{cycles}=2$ (``W-cycles'') and $N_{cycles}=3$.  We have measured
the residual reduction rate and the total solver time to solve to a
given accuracy ($10^{-10}$ in our tests) for a simple disk domain as
shown on Figure~\ref{fig:rate}.

\begin{figure*}
\includegraphics[width=0.45\textwidth]{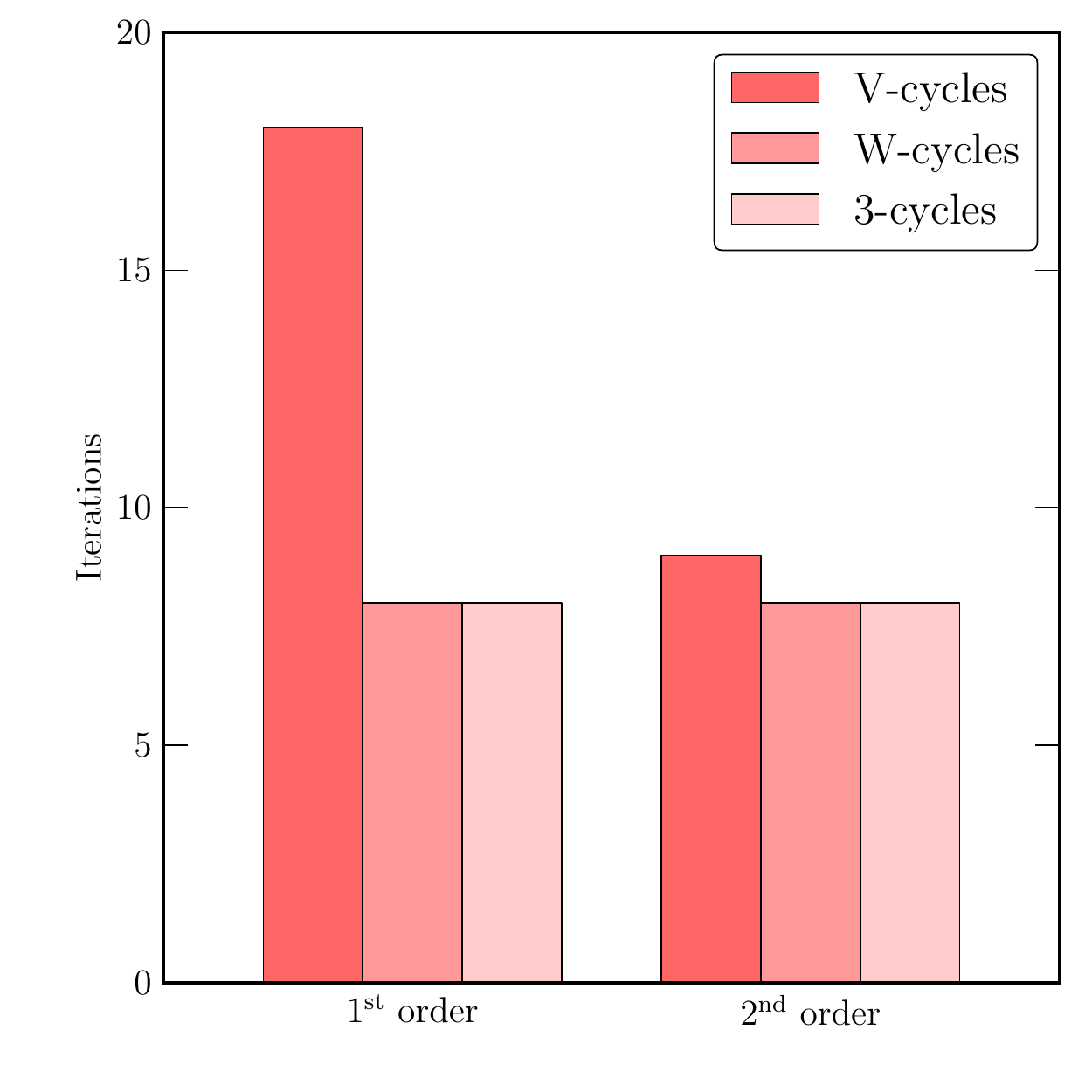}
\includegraphics[width=0.45\textwidth]{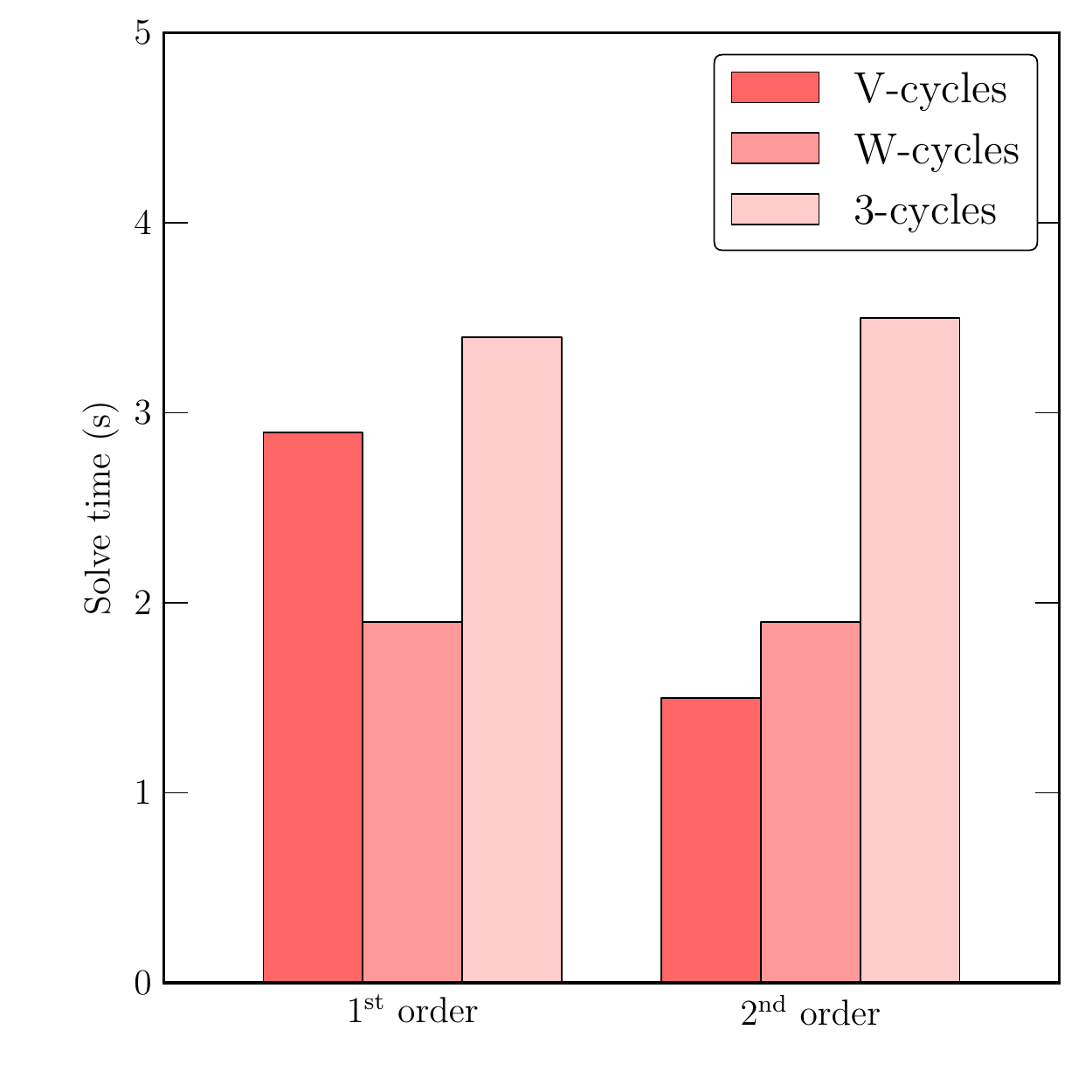}
\caption{Effect of the multigrid schedule on the solver, for a simple disk-shaped domain
  test case where second-order multigrid boundary reconstruction does converge.}
\label{fig:sched}
\end{figure*}

Typical results are presented on Figure~\ref{fig:sched}.  The first
conclusion is that V-cycles are very sensitive to the chosen boundary
reconstruction scheme. First-order multigrid boundary reconstruction is clearly detrimental
to the convergence rate, though it does ensure convergence of grids
with small islands.
Interestingly enough, schedules with $N_{cycles} \geq 2$ (like W-cycles) appear to be  insensitive to the
order of the boundary reconstruction scheme. This suggests that first-order multigrid boundary reconstruction used in
conjunction with W-cycles would ensure a convergence of the solver
as fast as the second-order scheme. Since W-cycles have more
costly iterations, we see in Figure~\ref{fig:sched} that this additional cost
translates however into a longer overall  time. 
For most astrophysical applications we have explored with the
RAMSES code, we have found that V-cycles
perform generally better than W-cycles. This can however depend on the
actual implementation of the solver, and the kind of grid geometries
arising in other applications.

\subsection{Parallel computing}

The progress of distributed memory architectures over vector
supercomputers has led to a regain of interest in iterative methods,
as direct global solvers such as FFT are particularly expensive in
terms of inter-process communications.  Iterative methods can often be
adapted to distributed memory architectures, with little modification,
and therefore remain simple to implement, while requiring limited
communications.

A broad class of parallelization techniques for physical problems
consists of splitting the computational domain into subregions, which
are each managed and updated by a dedicated computing core. Such
\emph{spatial domain decomposition} techniques rely on the ability to
update each CPU independently first, then address the couplings
between different domains.

\begin{figure} % {{{ Fig: buffer zones
\begin{center}
\beginpgfgraphicnamed{buffer-zones}
\begin{tikzpicture}[scale=0.8]

\newcommand{\MvU}{++(0, 1)}
\newcommand{\MvL}{++(-1, 0)}
\newcommand{\MvR}{++(1, 0)}

\newcommand{\cpuboundpath}{ (5,0) -- \MvU -- \MvL -- \MvU -- \MvL -- \MvU -- \MvR --
\MvU -- \MvR -- \MvU -- \MvR -- \MvU -- \MvU -- \MvL -- \MvU }

\newcommand{\bufApath}{ (4, 0) -- \MvL -- \MvU -- \MvL -- \MvU -- \MvU -- \MvU
-- \MvR -- \MvU -- \MvR -- \MvU  -- \MvU -- \MvU }

\newcommand{\bufBpath}{ (6, 0) -- \MvU -- \MvU -- \MvL -- \MvU -- \MvR -- \MvU -- \MvR -- \MvU -- \MvU -- \MvU -- \MvU -- \MvL}

\newcommand{\cpuAcol}{red!50}
\newcommand{\cpuBcol}{blue!50}
\newcommand{\bufAcol}{red!25}
\newcommand{\bufBcol}{blue!25}

% CPU1 active grids
\fill[\cpuAcol] (0, 0) -- \bufApath -- (0, 8) -- cycle;

% CPU2 active grids
\fill[\cpuBcol] (8, 0) -- \bufBpath -- (8, 8) -- cycle;

% CPU1 buffer region
\begin{scope}
  \clip (0, 0) -- \cpuboundpath -- (0, 8) -- cycle;
  \clip (8, 0) -- \bufApath -- (8, 8) -- cycle;
  \fill[\bufAcol] (0, 0) rectangle (8, 8);
\end{scope}

% CPU2 buffer region
\begin{scope}
  \clip (0, 0) -- \bufBpath -- (0, 8) -- cycle;
  \clip (8, 0) -- \cpuboundpath -- (8, 8) -- cycle;
  \fill[\bufBcol] (0, 0) rectangle (8, 8);
\end{scope}

% Draw the grid
\draw (0, 0) grid +(8, 8);

% Draw the CPU boundary
\draw[very thick] \cpuboundpath;
% Draw the text
\node[rectangle, inner sep=6pt, fill=\cpuAcol] at (1.5, 6.5) {\sc{CPU1}};
\node[rectangle, inner sep=6pt, fill=\cpuBcol] at (6.5, 2.5) {\sc{CPU2}};

\end{tikzpicture}
\endpgfgraphicnamed
\end{center}
\caption{Domain decomposition and buffer regions as used in RAMSES, in
particular for the Poisson solver. The thick black line marks the boundary of
the spatial domain decomposition between CPU 1 and CPU 2.  CPU 1 owns all the
red cells, while CPU 2 owns all the blue cells. In order to perform an update,
each CPU needs the values of the fields in the immediate exterior vicinity of
its domain (light red and light blue for CPUs 2 and 1 respectively). These
buffer cells are updated after each iteration of the various solvers using
inter-process communications.}
\label{fig:buffer}
\end{figure}
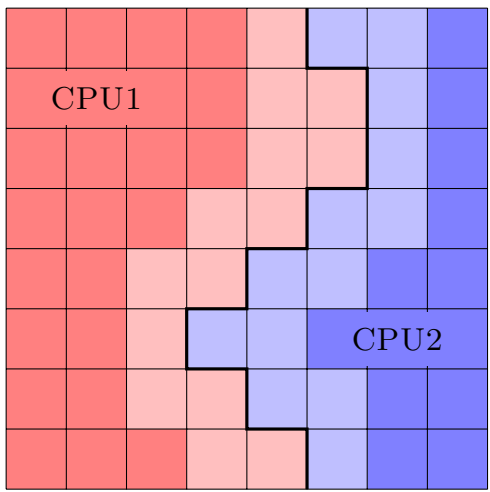

In RAMSES, this last step is implemented using buffer regions (see
Fig.~\ref{fig:buffer}). Each computing core manages its own cells, but
also possesses a local copy of cells from other neighboring CPUs which
are needed for local computation. These buffer cells need to be
updated after every iteration of the various iterative solvers used in
the code. The update operation is done by communicating the updated
values of the buffer cells from the CPUs which own them to the buffer
regions in other processors.
Therefore, any CPU only communicates with its direct neighbors, and the number of neighbors usually remains small.
Moreover, the number of buffer cells scales only as a surface term, limiting the transfer to computation volume ratio.
This is in contrast to the FFT, which requires a full transpose of the grid, and global all-to-all communications.
In our multigrid scheme, we need to communicate both the solution and the residual for each the buffer cell
between every Gauss-Seidel sweep, and also after each restriction or prolongation operation.

 % }}}

\begin{figure*}
\includegraphics[width=0.45\textwidth]{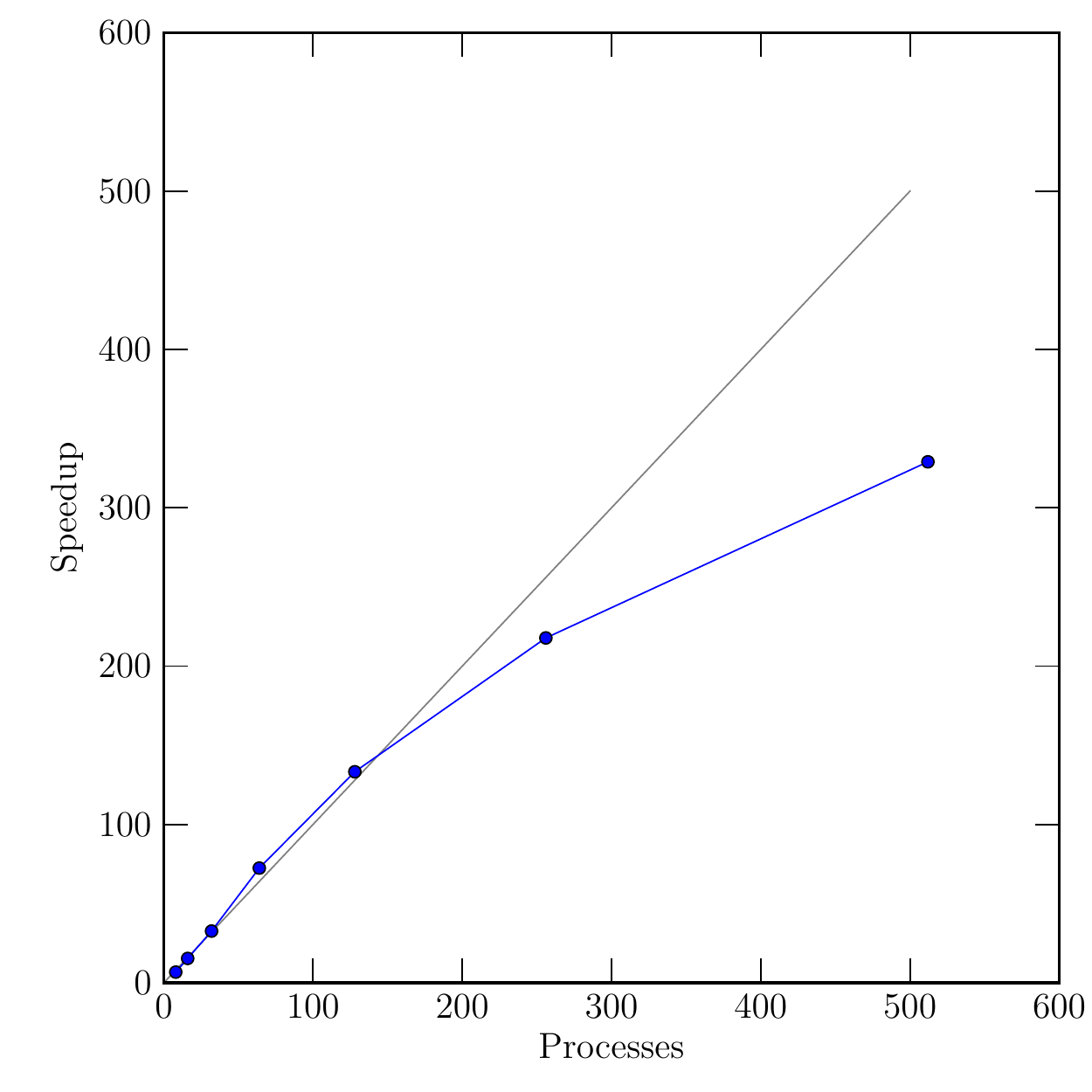}
\includegraphics[width=0.45\textwidth]{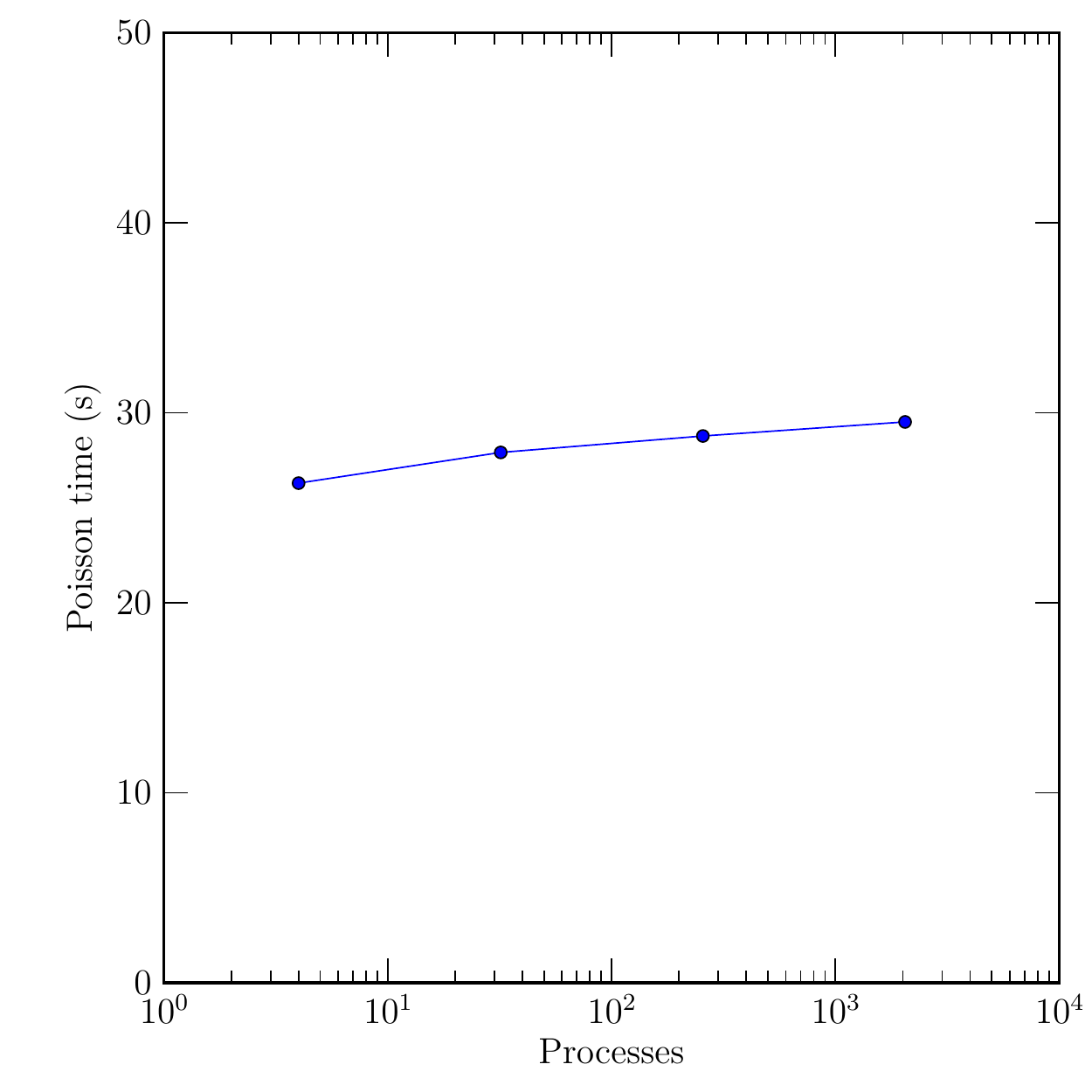}
\caption{Strong (left panel) and weak (right panel) scaling test of our multigrid Poisson solver for a
  $256^3$ Cartesian  grid.}
\label{fig:scaling}
\end{figure*}

We have performed weak and strong scaling timings of our multigrid Poisson
solver in RAMSES. The strong scaling test case is a simple $256^3$ cosmological
simulation without any refinement (Cartesian grid), starting at 4 processes up to 512
processes.  The weak scaling test scales from $256^3$ with 4 processes
to $2048^3$ with 2048 processes.  The test results are presented on
Figure~\ref{fig:scaling}. We see that our parallel efficiency degrades down to 50\% when 
we reach $32^3$ cells per processor. Beyond this limit, we spend more time communicating data than
updating the solution during each Gauss-Seidel sweep. We could improve our scaling by a factor of 2 by hiding
surface cells communications by inner cells computations. The weak scaling tests shows that if we keep the computational
load above $64^3$ cells per processor, the scaling is almost perfect. 
This rule of thumb applies also for more complex grid geometries, although load balancing can degrade significantly in case of
very deep adaptive time step strategies.

\section{Conclusion}

We have presented a simple and efficient multigrid algorithm for
solving the Poisson equation on irregular domains defined on a regular
Cartesian grid. This kind of problem frequently arises in AMR codes
using a one-way interface strategy, and the method is therefore of particular
interest for astrophysical applications with multiple adaptive time steps. 
Using a second-order reconstruction scheme for the boundaries of the multigrid levels, we have shown that our
multigrid scheme features optimal convergence properties. Since we use a multigrid hierarchy 
orthogonal to the actual AMR grid, our memory requirements are minimal. In case of particularly complex 
boundary conditions, we have observed that our scheme fails to converge, an issue previously identified as the ``small island'' problem.
We have designed a simple fix to this problem, degrading our multigrid boundary reconstruction accuracy to first-order.
We have implemented this new technique in the RAMSES code, using a simple algorithm to determine ``on the fly'' 
which reconstruction technique should be used.
We have shown the solver to be second-order accurate for the potential even in the presence of AMR levels,
while the global force accuracy is close to second-order accuracy.
Near level interfaces, the one-way interface scheme causes degradation of the force to first order,
but only in very localized regions of codimension one.
We have measured significant performance gains over
our standard conjugate gradient solver, especially in the case of cosmological ``zoom-in'' simulations, 
where large fully-refined domains are present in the AMR grid. This simple and efficient multigrid solver could also be used for 
incompressible flow solvers, in presence of multiphase fluids or complex embedded solid boundaries.

\noindent
{\bf Acknowledgments}

This work was granted access to the HPC resources of CCRT under the allocation 2009-SAP2191 
made by GENCI (Grand Equipement National de Calcul Intensif).

\vfill
\bibliography{cosmo}

\end{document}